\documentclass[aip,pof,reprint]{revtex4-1}

\usepackage{graphicx}
\usepackage{dcolumn}
\usepackage{bm}
\usepackage{CJK}
\usepackage[utf8]{inputenc}
\usepackage[T1]{fontenc}
\usepackage{mathptmx}
\usepackage{etoolbox}
\usepackage{amssymb}
\usepackage{amsmath}
\usepackage[modulo]{lineno}
\usepackage{natbib}
\usepackage{booktabs}
\usepackage[colorlinks=true,linkcolor=blue,citecolor=blue
]{hyperref}
\usepackage{multirow} 
\usepackage{etoolbox}
\usepackage{tikz}
\newcommand*{\circled}[1]{\lower.7ex\hbox{\tikz\draw (0pt, 0pt)%
circle (.5em) node {\makebox[1em][c]{\small #1}};}}
\robustify{\circled}
\makeatletter
\def\@email#1#2{%
\endgroup
\patchcmd{\titleblock@produce}
{\frontmatter@RRAPformat}
{\frontmatter@RRAPformat{\produce@RRAP{*#1\href{mailto:#2}{#2}}}\frontmatter@RRAPformat}
{}{}
}%
\makeatother
\begin{document}
\begin{CJK*}{UTF8}{gbsn} 

\title{Lock-in effect of over-tip shock waves and identification of the escaping vortex-shedding mode in pressure-driven tip leakage flow}


\author{Xiaolong Tang (唐小龙)}
\email{tangxl@shu.edu.cn}
\affiliation{Shanghai Institute of Applied Mathematics and Mechanics (SIAMM), Shanghai Key Laboratory of Mechanics in Energy Engineering, School of Mechanics and Engineering Science, Shanghai University, Shanghai 200444, China} 
\affiliation{Key Laboratory of Aerodynamic Noise Control, China Aerodynamics Research and Development Center, Mianyang Sichuan 621000, China}
\author{Eldad J. Avital}%
\affiliation{School of Engineering and Materials Science, Queen Mary University of London, London E1 4NS, UK
}%
\author{Xiaodong Li (李晓东)}
\email{lixd@buaa.edu.cn}
\affiliation{%
School of Energy and Power Engineering, Beihang University, Beijing 100191, China
}%
\author{Fariborz Motallebi}%
\affiliation{School of Engineering and Materials Science, Queen Mary University of London, London E1 4NS, UK
}%
\author{Zainab J. Saleh}
\affiliation{School of Engineering and the Environment, Kingston University London, London KT1 2EE, UK}


\date{\today}

\begin{abstract}
Time-resolved schlieren visualization is used to investigate the unsteady flow structures of tip leakage flows in the clearance region. A common generic blade tip model is created and tested in a wind tunnel under operating conditions ranging from low-subsonic to transonic. A multi-cutoff superposition technique is developed to achieve better flow visualization. Quantitative image processing is performed to extract the flow structures and the instability modes. Additional numerical simulations are performed to help classify the observed flow structures. Unsteady flow structures such as over-tip shock oscillation, shear-layer flapping, and vortex shedding are revealed by Fourier analysis and dynamic mode decomposition. The results show that, under subsonic conditions, the trigger position of the shear layer instability is monotonically delayed as the blade loading increases; however, this pattern is reversed under transonic conditions. This implies that flow compressibility, flow acceleration, and the oscillation of over-tip shock waves are critical factors related to tip flow instabilities. The over-tip shock waves are observed to be locked-in by frequency and position with the shear-layer flapping mode. An intermittent flow mode, termed the escaping vortex-shedding mode, is also observed. These flow structures are key factors in the control of tip leakage flows. Based on the observed flow dynamics, a schematic drawing of tip leakage flow structures and related motions is proposed. Finally, an experimental dataset is obtained for the validation of future numerical simulations. 
\end{abstract}


\maketitle 

\section*{Nomenclature}

\begin{tabbing}
	xxxxxxxx \= xxxxxxxxxxxxxxxxxxxxxxxxxxxxxxx\= xxxxxxxxxx\= xx \kill
	$a$ \> passing height of schlieren image \\
	$b_i$ \> DMD mode amplitude \\
	$c$ \> tip clearance \\
	DMD \> dynamic mode decomposition \\
	$f$ \> frequency \\
	$f_m$ \> focal length of schlieren mirror \\
	$f_s$\> sample rate\\
	$g$ \> grayscale of images\\
	$h$ \> height of slit light source \\
	$h_b$ \> height of separation bubble \\
	$K,K_{sv},K_{sh}$ \> constants of schlieren system\\
	$l_b$ \> length of separation bubble \\
	$l_{max(h_b)}$ \>position of maximum bubble height \\
	$L$ \> length of schlieren extent \\
	$n_c\times n_w$ \> grid size in the clearance \\
	$n_p$\> number of images\\
	$P_{out}$ \> static pressure at $x=350$ mm \\
	$P^*$ \> atmospheric pressure\\
	$\Delta P$ \> $=P^*-P_{out}$, pressure difference \\
	$r_p$ \> $=\Delta P/P^*$, ratio of pressure difference \\
	$\bar{P}^{*}_{ref}$ \> total pressure at $x=-100$ mm\\
	$\bar{P}^*(x)$ \> averaged total pressure at $x$ \\
	$Re_{c}$ \> Reynolds number by clearance \\
	$Re_{tunnel}$ \>Reynolds number by tunnel height\\
	$St_c$ \> Strouhal number by clearance \\
	$T^*$ \> total temperature \\
	T1-T4 \> four time labels \\
	$v_x,v_y$ \> velocities \\
	$w$ \> width of model\\
	$x,y$ \> coordinates \\
	$\alpha$ \> absolute flow angle \\
	$\bar{\alpha}$ \> mass-weighted average of $\alpha$\\
	$\gamma(x)$ \> normalized grayscale at $x$ \\
	$\delta_i$ \> DMD mode growth rate \\
	$\epsilon(x)$ \> total pressure loss at $x$ \\
	$\eta$ \> schlieren cutoff ratio\\
	$\kappa$ \> Gladstone--Dale coefficient \\
	$\rho$ \> density\\
	$\rho_x,\rho_y$ \> normalized density gradient\\
	$\rho^*_x,\rho^*_y$ \> density gradient\\
	$\mathbf{\phi}_i$ \> DMD eigenvector \\
	$\phi_x,\phi_y$ \> scaled density gradient \\
	\\
	\textit{Subscripts:}\\
	a1--a4 \> four DMD modes from vertical left cutoff\\
	b1--b4 \> four DMD modes from horizontal down cutoff\\
	h1--h4 \> four Fourier modes from horizontal down cutoff\\
	v1--v4 \> four Fourier modes from vertical left cutoff\\
\end{tabbing}
\section{Introduction}\label{s:intro}
\begin{figure}[!b]
\begin{minipage}[t]{1\linewidth}
\includegraphics[width=3.2in]{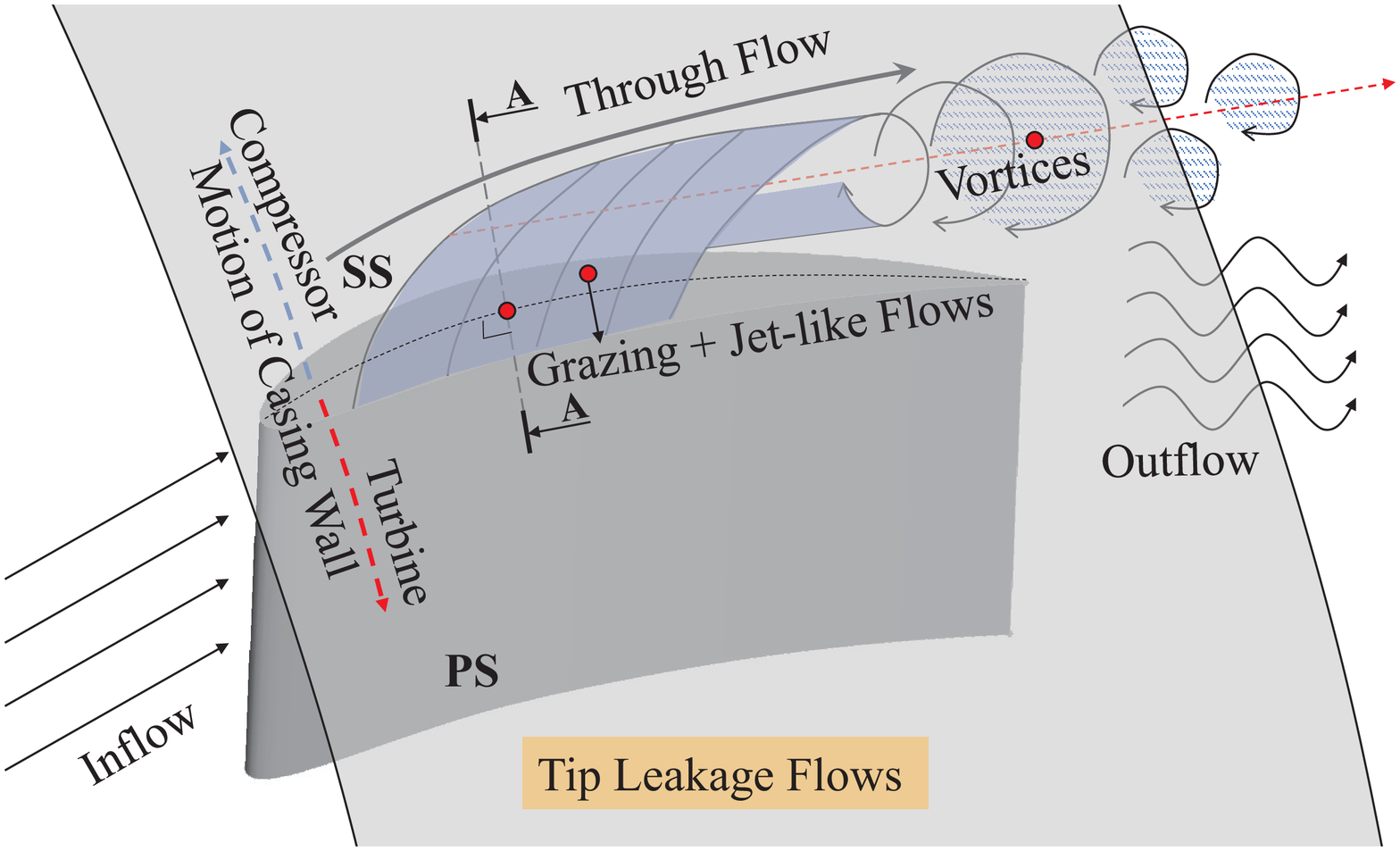}\\{\small {(a) Overall view.}}
\end{minipage}
\begin{minipage}[t]{1\linewidth}
\centering
\includegraphics[width=2.2in]{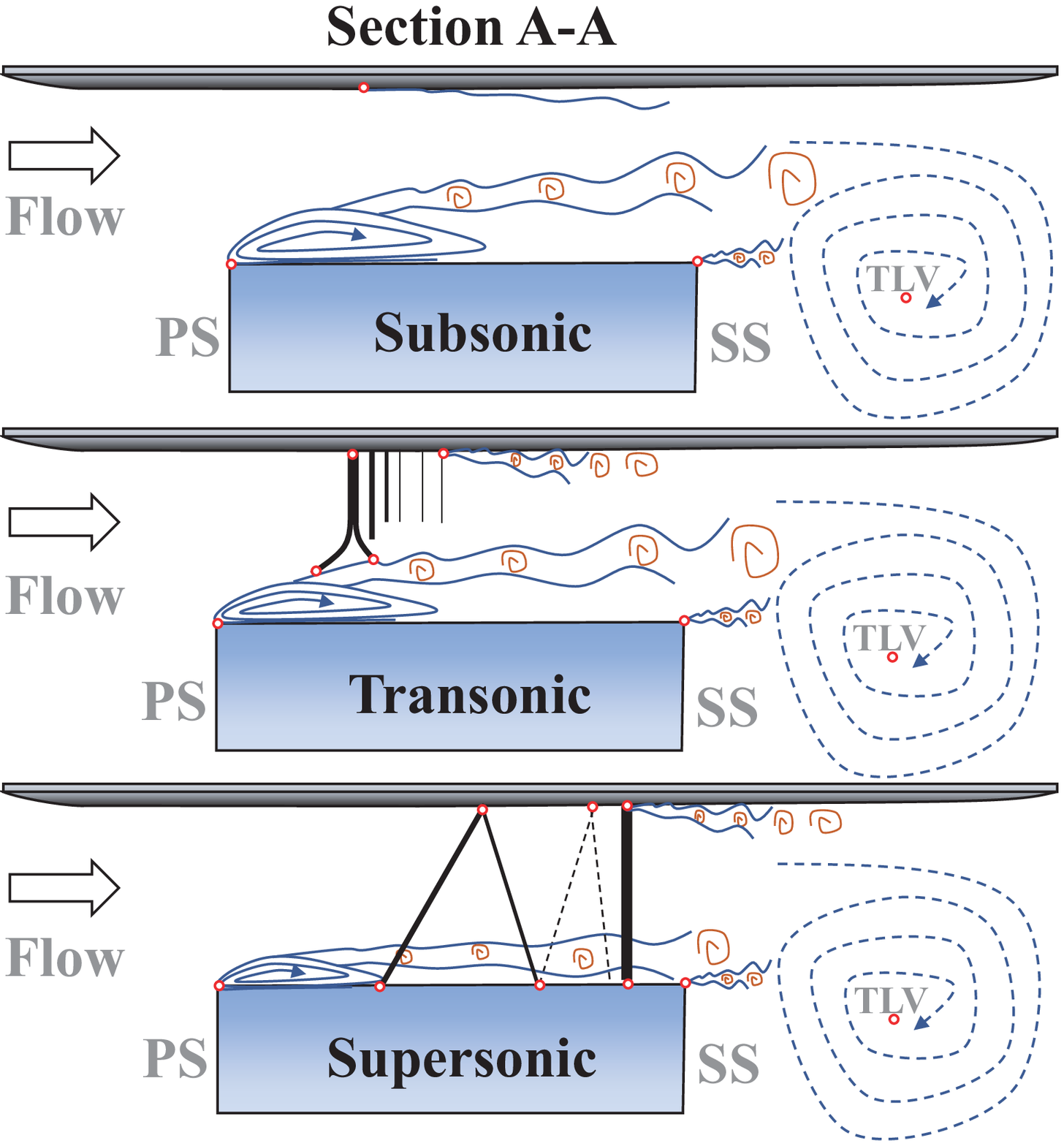}\\
{\small {(b)} Tip leakage flows along section A-A.}
\end{minipage}
\caption{\label{g:tip_flows_tips} Schematic view of leakage flows over the blade tip: (a) overall view; (b) view along section A-A. Three operating conditions with different clearance-flow structures are displayed for section A-A. These flow structures, such as separation bubbles, vortex shedding, and shock waves and their interactions, are verified in the following context. }
\end{figure}

\setlength{\parskip}{0ex}
In most cases, rotor-blade tip clearance is inevitable in turbomachinery. This leads to tip leakage flows (TLFs), as shown in Fig.~\ref{g:tip_flows_tips}, and related flow--flow or flow--blade interactions, such as the rolling-up of tip leakage vortices (TLVs). TLFs are profoundly important because of their significant influence on efficiency loss \cite{you2007-jfm,coull2015,deveaux2020,wang2022-pof}, heat transfer and corrosion fatigue \cite{mischo2008,zhang2011-jt,wheeler2011,wheeler2016}, flow instability \cite{hewkin-smith2017,tian2018-exif,wanghao2020-ast}, noise generation \cite{khorrami2002,pogorelov2015,bizjan2016,zhang2020-ast,palleja-cabre2022} and cavitation \cite{li2021-pof}. The loss introduced by TLFs constitutes about one-third of the total stage loss \cite{denton1993}, and about half of this loss occurs in the gap region \cite{bindon1989}. Severe blade tip heat loads are the main cause of turbine blade degradation. This is closely related to the state of tip heat transfer, which is very sensitive to the tip flow structures. Research has shown that the blade-tip heat load within a high-pressure turbine varies by up to 50\% when switching from sub- to supersonic operating conditions \cite{wheeler2011}. Under high loading conditions, TLFs may lead to strong flow instability and rotating stall in compressors \cite{hewkin-smith2017}. TLFs and TLVs are also the main flow irregularities in blade stages. These irregularities interact with the blade trailing edge and adjacent blades to generate significant noise \cite{milavec2015,jacob2016,luo2020-ast}.

Previous investigations have largely focused on reducing the loss and heat load, or the unsteadiness of tip flow structures outside the clearance region \cite{he2021-pof,palleja-cabre2022,hou2022-pof,babu2022-pof}. However, the origin of such flow structures is the clearance, the unsteadiness of which serves as the initial trigger for the instability inside and outside the tip region. Limited by the narrow space in the tip region, it remains difficult to obtain detailed flow measurements in the clearance region, although Jacob et al. \cite{jacob2016} and Feng et al. \cite{feng2021-pof} have shed some light on this topic. This difficulty has been partially overcome by high-fidelity numerical simulations \cite{wheeler2016,shang2021-pof}, although further validation using experimental data is essential. Additionally, TLF structures and the related motions under transonic conditions are yet to be clarified.

Several flow field visualization techniques, such as oil-film \cite{lee2010-exif,lee2016-ijhff}, schlieren \cite{saleh2019,feng2021-pof}, laser Doppler velocimetry, \cite{mailach2008,deveaux2020} and particle image velocimetry \cite{jacob2016}, have been used to study TLFs. The effects of clearance size and tip contouring on the TLF-induced unsteadiness patterns, turbulent kinetic energy, and passage blockage have subsequently been reported. Fischer et al. \cite{fischer2013} proposed a frequency-modulated Doppler global velocimetry method to cope with the limited measuring space and high unsteadiness in the tip region. Recently, a time-resolved schlieren visualization technique was used by Feng et al. \cite{feng2021-pof} to investigate the rapid establishment of the in-gap TLF under transonic operating conditions. Their results showed that transonic TLFs could be established within 0.65 s through five stages. Time-resolved schlieren visualization can also be used for quantitative analysis through Fourier transforms and dynamic mode decomposition (DMD). These methods have been widely used to analyze wakes \cite{shi2022-pof} and confined gap flows \cite{karthick2021-pof}. Rao and Karthick \cite{rao2019} have discussed the influence of image parameters on the DMD of time-resolved schlieren images. These studies provide effective methods for detailed investigations of the TLFs generated by various tip models.

The TLF in the clearance region is composed of three subflows: (1) pressure-driven potential flow; (2) viscous flow due to the blade tip and the casing wall boundaries; and (3) viscous flow due to the rotating casing wall. The contributions from the first two subflows can be captured by two-dimensional blade-tip simplification with a stationary casing wall \cite{wheeler2011,wheeler2016,feng2021}. This is because, once the pressure difference has been determined by the main flow in the blade passage, the in-gap TLFs become relatively independent \cite{chen1991}. The stationary casing-wall assumption is a compromise; fortunately, many critical features of compressor TLFs are unaffected by a moving casing wall \cite{storer1994,wang2004-aiaaj}. This is valid in the case of no strong inlet distortions \cite{coull2015}. Because of the different motion directions, as shown in Fig.~\ref{g:tip_flows_tips}, the influence of wall motion is strong in turbines \cite{yaras1992a} and weak in compressors \cite{nikolos1995}. Considering the different effects of a moving casing wall on compressors and turbines, some critical common flow features can be observed in both compressors and turbines with a stationary casing wall. 

The main objective of this study is to investigate the steady and unsteady structures of TLFs under multiple operating conditions, resulting in a better understanding of the mechanisms of TLFs. The emphasis is transonic conditions, because the TLFs become more unsteady under high loading conditions \cite{furukawa1999}, when the flow field through the rotor stage can be strongly influenced by the unsteady motion of TLFs \cite{hah2008}. This goal is achieved by wind-tunnel testing with time-resolved schlieren visualization and image processing, and is supported by additional numerical simulations. Multiple schlieren cutoffs are applied, which enables the quantitative visualization of flow structures in both the horizontal and vertical directions. The experimental and numerical configurations are described in Sec.~\ref{s:exp_setup}. In Secs.~\ref{s:results1}--\ref{s:results3}, postprocessing is applied and the results are analyzed and discussed. Finally, the conclusions to this study are presented in Sec.~\ref{s:conclude}.





\section{Experimental and numerical configurations}\label{s:exp_setup}
\subsection{Simplified tip models}
Two-dimensional simplified blade tip models with a stationary casing wall are used. The model profile, which is illustrated by section A-A in Fig.~\ref{g:tip_flows_tips}, is representative of the transverse section of the tip part of rotor blades. This enables detailed observations of the flow structures in the clearance region. However, we do not focus on a specific position along the blade chord, because the position of the strongest TLF varies along the chord in different applications. This position tends to be near the leading edge in compressors/fans \cite{suder1996}, at the mid-chord for propeller rotors \cite{jang2001-jfe1}, and near the mid-to-trailing-edge chord for turbines \cite{azad2000}. Several similar simplifications have been applied in the literature \cite{nikolos1995,wheeler2016,fang2019,saleh2019,feng2021-pof}. 

\begin{figure}[!htb]
\centering
\includegraphics[width=3.4in]{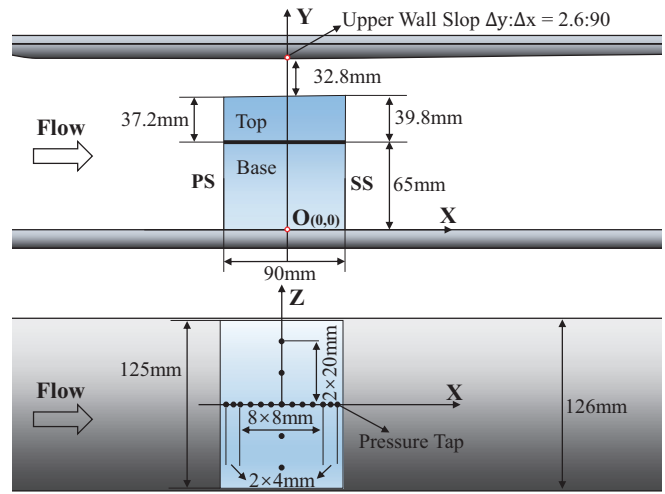}
\caption{\label{g:top_dimension}Schematic view of the model mounted in the wind tunnel, with a frontal view (upper) and top view (lower). The black dots represent the pressure tapping holes. The models have a shared base and a replaceable top mounted on the base. The 13 taps along the X-axis are equally distributed at 8 mm intervals, except for the first and the last three, which have smaller intervals of 4 mm. The five taps along the Z-axis are equally distributed at intervals of 20 mm.}
\end{figure}

As shown in Fig.~\ref{g:top_dimension}, the simplified two-dimensional models are stretched along the depth direction and enlarged to fit into the wind tunnel, which has a rectangular working section.
The overall model height at the pressure side (PS) is 102.2 mm, which is composed of a 65 mm base and a 37.2 mm top. On the suction side (SS), the top measures 39.8 mm, giving a model height of 104.8 mm at the SS edge (with a height difference of 2.6 mm). This height difference is conformal with the upper wall of the wind tunnel (see subsection \ref{s:wind_tunnel}) to maintain a constant tip clearance. The model width and the tip clearance are 90 mm and 32.8 mm, respectively. This yields a clearance to blade-thickness ratio of 36.44\% (2--5\% chord), which is within the typical clearance ratio range in fans, compressors, and turbine blades \cite{bindon1989,bringhenti2008}. Moore et al. \cite{moore1992} pointed out that the influential area of the core part is limited to twice the tip clearance for transonic tip leakage flows. This value is taken as the criterion for avoiding severe tunnel blockage. Thus, the present tunnel height to clearance ratio is set to approximately 4.2 at the clearance exit, which is more than twice the core-part size. Seventeen pressure taps are placed on the top surface. Details of the pressure taps are illustrated in Fig.~\ref{g:top_dimension}. The five taps along the Z-axis are used to monitor the spanwise effect to ensure a two-dimensional flow pattern. 

\subsection{Wind tunnel}\label{s:wind_tunnel}
\begin{figure}[!htb]
\begin{minipage}[t]{1.0\linewidth}
\centering
\includegraphics[width=3.4in]{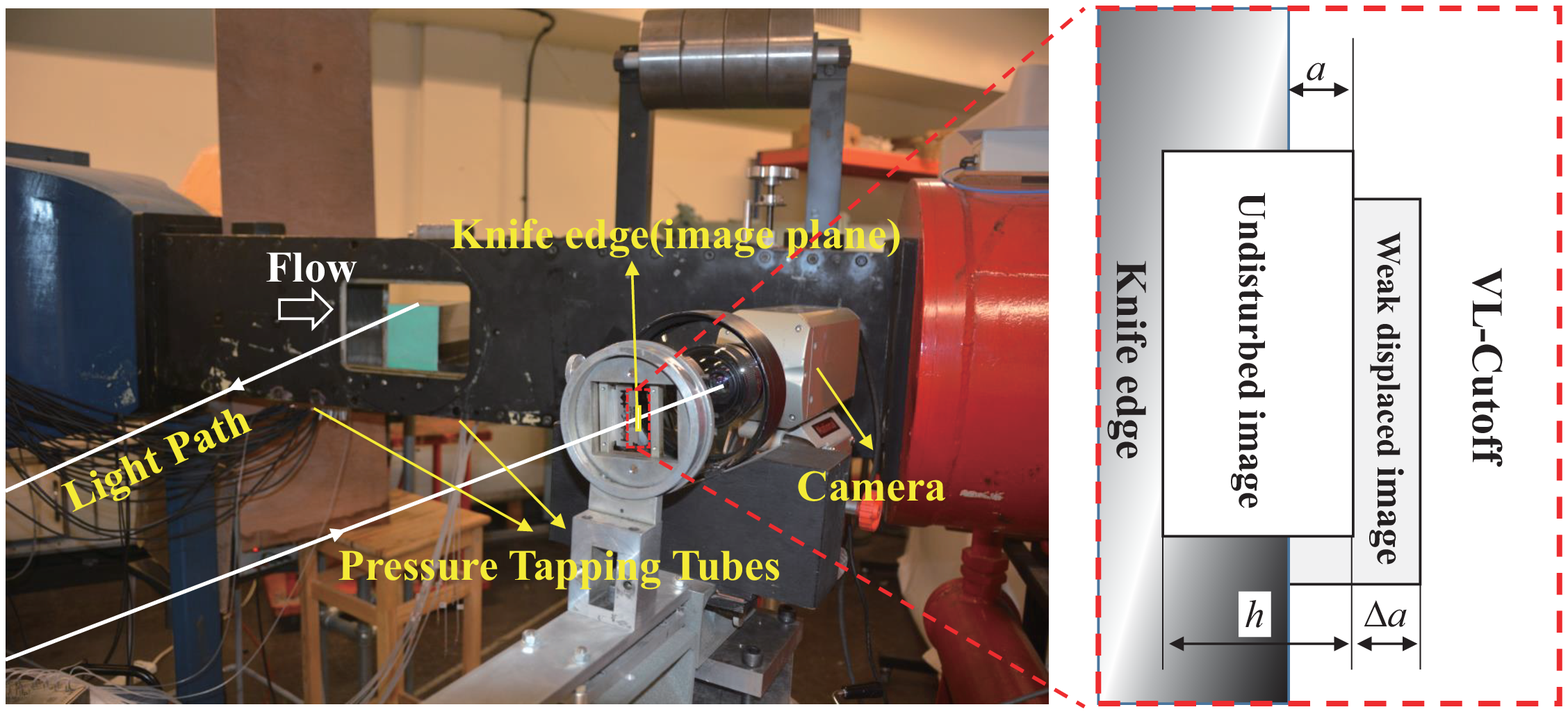}\\{\small {(a) Photo of the wind tunnel.}}
\end{minipage}
\begin{minipage}[t]{1.0\linewidth}
\centering
\includegraphics[width=3.4in]{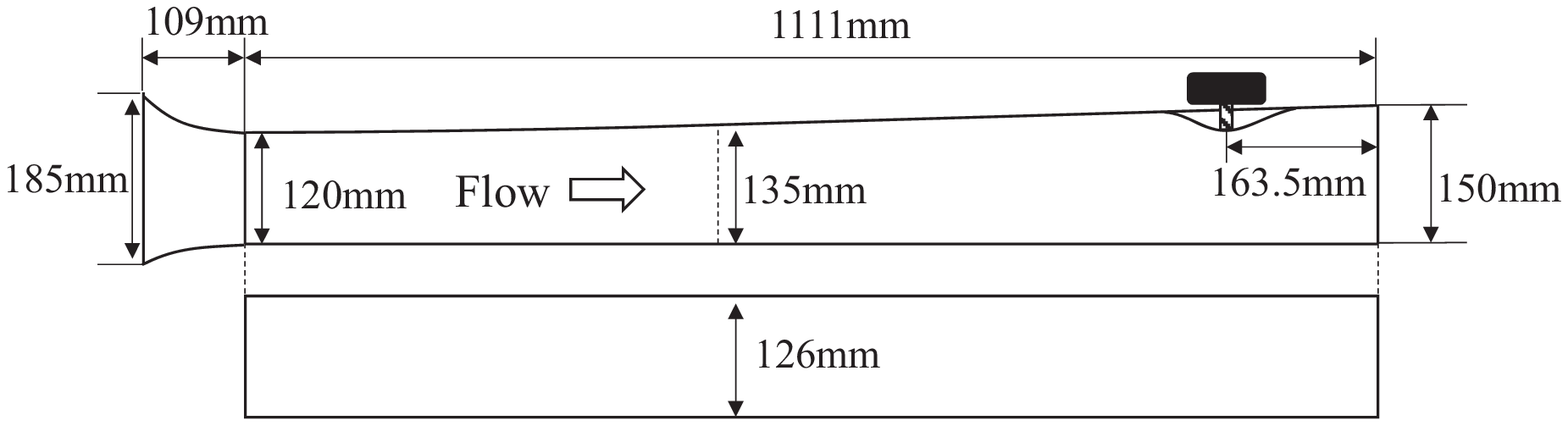}\\{\small {(b) Profile of the wind tunnel.}}
\end{minipage}
\caption{\label{g:photos} Wind tunnel: (a) photo of the closed-circuit transonic wind tunnel with the simplified model and part of the schlieren system (see subsection \ref{subs:schlieren}); (b) frontal- and top-view profiles of the wind tunnel.}
\end{figure}

The transonic wind tunnel \cite{shahneh2009,saleh2019} of the Whitehead Aeronautical Laboratory at Queen Mary, University of London, was used for the investigations. Figure \ref{g:photos} shows a photo of the wind tunnel, part of the schlieren system, and the dimensions of the working section. This is a closed-circuit transonic wind tunnel. The working section used in this study has a width of 126 mm and a height of 135 mm. [The upper wall of the tunnel is slightly divergent in this section, with a slope ratio of about 2.89\% (2.6:90), so the model top surface is machined to have a small slope and maintain a constant tip clearance.] The gap between the tip and the upper (casing) wall ranges from 31.3\% to 32.1\% of the wind tunnel working section height. Steady operation of the wind tunnel is ensured under this clearance ratio. The nominal working Mach number without models ranges from 0--1.4. The small gap between the quartz screen and the model is sealed with a sponge. The small gaps between the tunnel floor and the model parts were sealed with aluminum tape.

The driving pressure, termed the pressure demand (PD), is the pressure supply of the wind tunnel driving system. It ranges from 0--827.37 kPa (0--120 psi). The inlet total pressure $P^*$ is measured in the settling chamber immediately before the test section, which has a constant atmospheric pressure. The outlet static pressure $P_{out}$ is measured at the tunnel floor, 350 mm downstream of the model center ($x=0$ mm), where the static pressure becomes relatively uniform. Consequently, the ratio of pressure difference imposed on the model is defined as $r_p=(P^*-P_{out})/P^*$, which ranges from 0.0--0.297. The wind tunnel runs stably with model when PD $\le$ 827.37 kPa (PD $\le$ 120 psi, $r_p\le0.297$). When using the simplified model with the maximum PD (labeled as PD120), the Reynolds number of the tunnel [calculated using the tunnel height] is approximately $Re_{tunnel}=3.9\times10^{5}$, and the Reynolds number calculated at the exit of the tip gap is approximately $Re_{c}=4.1\times10^{5}$. The averaged Mach number in the tunnel is approximately 0.12, and the Mach number at the center of the clearance exit is approximately 0.84. The highest Mach number near the inlet of the gap is approximately 1.2. As listed in Table \ref{t:pressure demand}, the pressure ratios are obtained by averaging over multiple runs. These data show the overall performance of the wind tunnel running with the blade tip model. The environmental pressure and temperature are relatively stable during the tests. The total temperature varies from 290.8--293.3 K. The inlet total pressure is calibrated to be $P^*=$101.33 kPa at a total temperature of $T^*=288.15$ K.
\begin{table}[!htb]
\small
\caption{List of the pressure demand (PD, driving pressure) and the static-total pressure ratio in the wind tunnel with the model, averaged over 2--4 runs. A linear relation between $r_p$ and PD is assured.}
\label{t:pressure demand}
\centering
\begin{tabular}{c|c|c|c|c|c|c}
\hline
PD (psi) & 10 & 20 & 30 & 40 & 50 & 60  \\
\hline
PD (kPa) & 68.95 &137.90 & 206.84 &275.79 & 344.74&413.69 \\
\hline
$r_p$ & 0.035&0.063&0.091&0.115&0.141&0.166 \\
\hline
PD (psi) & 70 & 80 & 90 & 100 & 110 & 120 \\
\hline
PD (kPa) &482.65 & 551.58 & 620.53 & 689.48 & 758.42 & 827.37 \\
\hline
$r_p$ &0.191&0.214&0.238&0.263&0.281&0.297 \\
\hline
\end{tabular}
\end{table}

\subsection{Time-resolved schlieren visualization system}\label{subs:schlieren}
\begin{figure*}[!htb]
\centering
\includegraphics[width=5.5in]{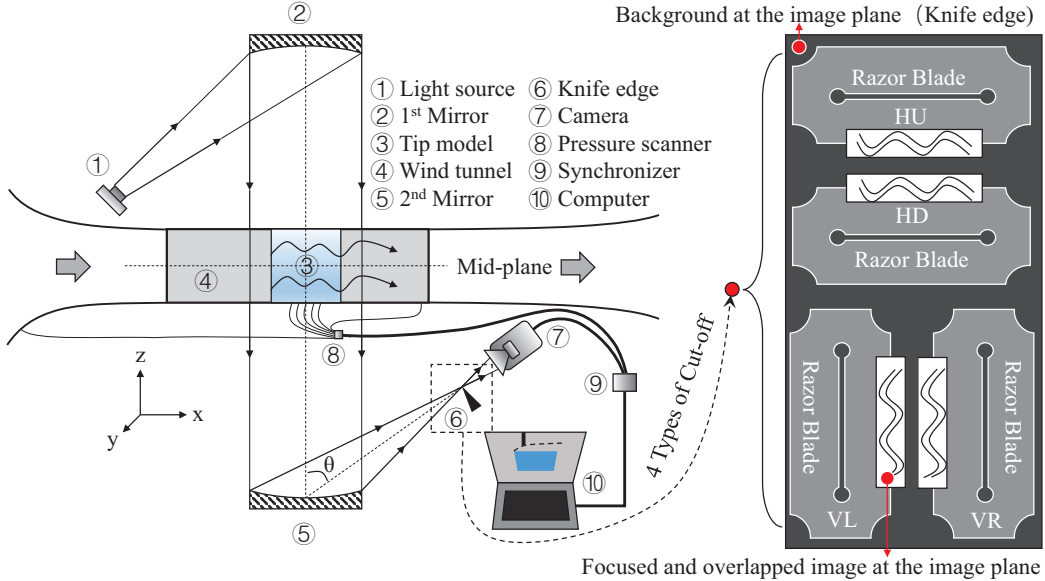}
\caption{\label{g:schlieren_system}Schematic drawing of the symmetric $z$-type schlieren and pressure testing system.}
\end{figure*}

A symmetric $z$-type schlieren system was applied in this study (see Fig.~\ref{g:schlieren_system}). Thus, the comma error  of schlieren systems is avoided by the symmetry. The component parameters of this system are listed in Table \ref{t:schlieren_para}. The angular deflection of the light ray is linearly proportional to the grayscale captured by a camera that saves the image without gamma correction \cite{gamma-correction}. Thus, the grayscale measured by the camera, $(\phi_x,\phi_y)$, has a linear relationship with the flow density gradient, i.e., $\phi_x=K\partial \rho/\partial x$, $\phi_y=K\partial \rho/\partial y$, where $K$ is a constant determined by the light-source illuminance and parameters of the camera. Assuming the environment refractive index $n_0\approx 1$, the flow density gradient is expressed as
\begin{equation}\label{eq:density_gradient}
\frac{\partial \rho (x)}{\partial x}\cong \frac{h[ \gamma(x)-a/h]}{\kappa Lf_m},
\end{equation}
where $\gamma(x)$ is the normalized grayscale. As shown in Fig.~\ref{g:photos}, $h=0.3$ mm is the height of the slit light source and $a$ is the passing height of the image at the knife-edge plane. Thus, $a/h$ represents the cutoff ratio. Generally, the value of $a$ varies from $0.2h$--$0.8h$ depending on the camera's need for proper exposure and the balance between the schlieren sensitivity and the measuring range. The length of the schlieren extent is $L=126$~mm (width of the wind tunnel). The focal length of the mirror is $f_m=1828.8$~mm, and the Gladstone--Dale coefficient is $\kappa \approx0.23$~cm$^3\cdot$ g$^{-1}$.
\begin{table}[!htb]
\small
\centering
\caption{Parameters of the schlieren components.}\label{t:schlieren_para}
\begin{tabular}{c|c|c|c|c}
\hline
Light Source& \multicolumn{2}{c|}{KEYMED slit source} & Slit height & 0.3 mm \\
\hline
Mirrors & Focus & 1828.8 mm & Diameter & 203.2 mm \\
\hline
Knife edge & \multicolumn{2}{c|}{Razor blade} & Extent Length& 126 mm\\
\hline
Lens & \multicolumn{4}{|c}{55-80 mm}\\
\hline
\multirow{3}{*}{Camera}& Model & Phantom V4.3 & Sample rate & 8510 Hz \\
\hline
&Exposure & $10~\mu$s & Sample time & 1.888 s\\
\hline
&\multicolumn{2}{c|}{Resolution (pixels)}& $512\times128$ \\
\hline
\end{tabular}
\end{table}

The application of a cutoff is essential for schlieren measurements. The cutoff type denotes the direction from which the slit light at the knife-edge plane is partly blocked, as shown in Fig.~\ref{g:schlieren_system}. Four types of schlieren cutoffs are used. They are the vertical-left (VL), vertical-right (VR), horizontal-down (HD), and horizontal-up (HU) cutoffs, and the slit light source and the edge of the razor blade are placed vertically or horizontally according to the vertical- or horizontal-cutoff types. Multiple cutoff directions of a schlieren system are necessary for quantitative measurements, though only one cutoff direction is used \cite{feng2021-pof,whalen2021} when forming a qualitative visualization of the flow. This is important because the sensitivity and the effective measuring range of a schlieren system are controversial \cite{settles2001}. To extend the effective measuring range, a superposition technique is applied. The cutoff ratio $\eta=(h-a)/h$ varies within a range of $[20\%,80\%]$ in the experiments. The superposition of two images measured by two cutoffs in opposite directions is used to achieve a larger measuring range.

\begin{table}[!htb]
\small
\caption{Relationships between cutoff type and flow structures that can be revealed by schlieren patterns in compressible flows (as verified in the following discussion).}
\label{t:relation_flow_cutoff}
\centering
\begin{tabular}{ccl}
\toprule
Cutoff & ~~Good at capturing~~ & Possible flow structures revealed \\
\midrule
VL & $\partial \rho/\partial x>0$ & Shocks, vortex--shock interaction\\
\midrule
VR & $\partial \rho/\partial x<0$ & Separation bubble\\
\midrule
HD & $\partial \rho/\partial y>0$ & Shear layer, shear--wall interaction\\
\midrule
HU & $\partial \rho/\partial y<0$ & Shear layer, shear--wall interaction\\
\bottomrule
\end{tabular}
\end{table}
The flow patterns revealed by schlieren images are essentially the flow compression or expansion in the $x$- or $y$-direction. Flow structures such as shock waves, separation bubbles, expansion waves, and shear layers are accompanied by strong density gradients. Thus, the schlieren patterns in turn reveal such flow structures. According to preliminary knowledge about TLFs \cite{moore1992}, the mappings between the schlieren cutoff type and the revealed flow structures are listed in Table \ref{t:relation_flow_cutoff}. The flow structures that exhibit an abrupt density increase near the upper wall can be identified as shock waves. Abrupt density changes near the pressure side edge are identified as the boundaries of separation bubbles. Strong density gradients in the $y$-direction denote the boundary of the shear layer. The periodic shedding of negative--positive density gradient patches is identified as vortex shedding, and these negative--positive density gradient zones are identified as vortices. A continuous increase or decrease in density is identified as a compression or expansion region.

\subsection{Numerical setup}
\begin{figure}[!htb]
\centering
\includegraphics[width=3.in]{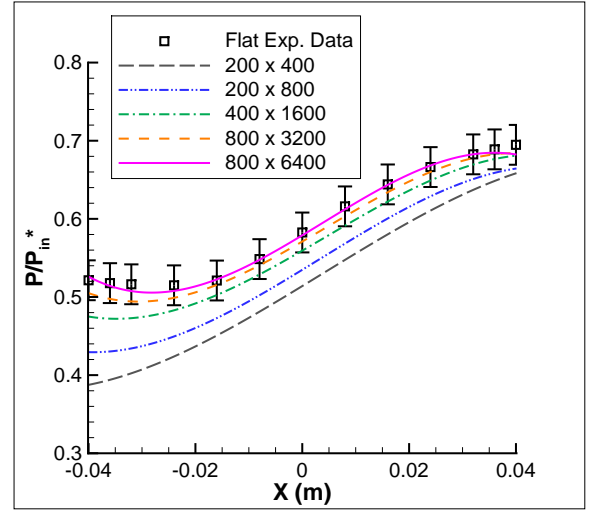} 
\caption{\label{g:mesh_sensitivity}Comparisons of the predicted and measured static pressure distributions (time-averaged) at the tip surface at $r_p=0.297$ (PD120). Five sets of clearance grids were tested. Error bars were created based on the accuracy level of the pressure scanner.}
\end{figure}

\begin{figure}[!htb]
\begin{minipage}[b]{1\linewidth}
\includegraphics[width=3in]{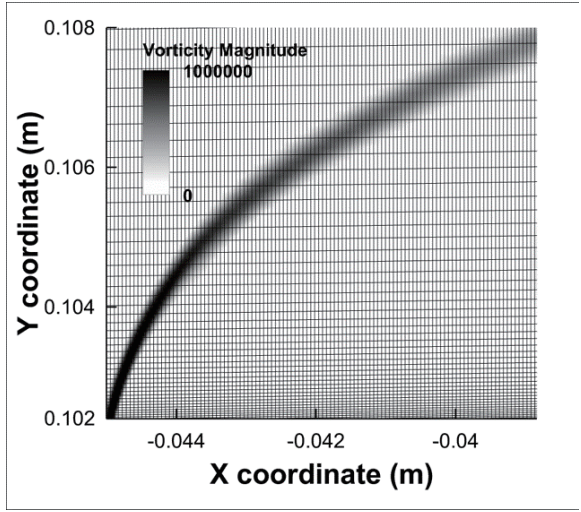}
{\\\small{(a) Vorticity magnitude at $r_p=0.166$ (PD80).}}
\end{minipage}
\hfill
\begin{minipage}[b]{1\linewidth}
\includegraphics[width=3in]{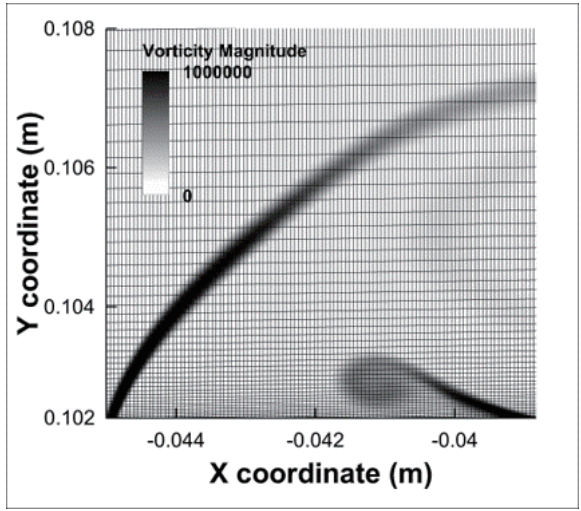}
{\\\small{(b) Vorticity magnitude at $r_p=0.297$ (PD120).}}
\end{minipage}
\caption{\label{g:mesh_vorticity} Computational mesh in the start region of the PS-side separation bubble, showing every fourth grid line. Instantaneous vorticity magnitude is displayed: (a) vorticity magnitude at $r_p=0.166$ (PD80); (b) vorticity magnitude at $r_p=0.297$ (PD120).}
\end{figure}

Numerical simulations were performed to qualitatively verify the flow structures observed by the schlieren system and to obtain additional flow information. Two-dimensional IDDES  simulations were applied using ANSYS Fluent. The results of grid sensitivity and flow validation tests are displayed in Fig.~\ref{g:mesh_sensitivity}. The time-averaged pressure distribution is in good agreement with the test data when the clearance grid is finer than $n_c \times n_w=800\times 3200$. Thus, the mesh in the gap region is set to $n_c \times n_w= 800\times 3200$. Mesh details in the start region of the PS-side separation bubble are illustrated in Fig.~\ref{g:mesh_vorticity}. The instantaneous vorticity magnitude is shown as an indicator of the boundary of the separation bubble. The final mesh has a dimensionless scale of $Y^+ \approx 1$ and $X^+ \approx 28$. The computational domain extends from $x=-0.2$~m to $x=0.5$~m, where the model is centered at $x=0$~m (see Fig.~\ref{g:top_dimension}). Thus, there are a total of about 4.4 million grid cells.
\begin{figure*}[!htb]
	\centering
	\begin{minipage}[t]{0.48\linewidth}
		\includegraphics[width=2.8in]{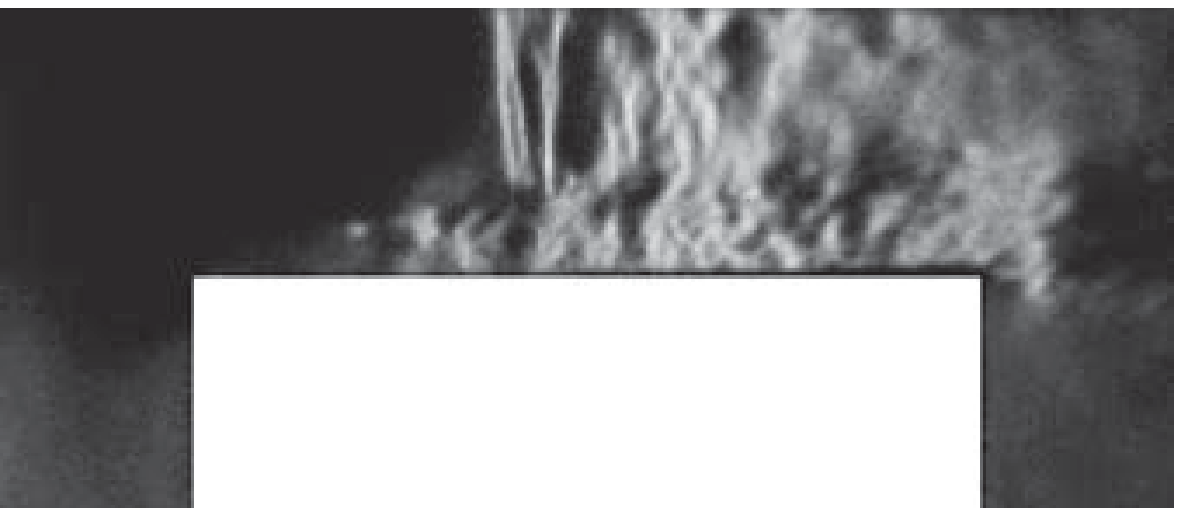} 
		{\\\small(a) VL-cutoff, unsteady.}
	\end{minipage}
	\begin{minipage}[t]{0.48\linewidth}
		\includegraphics[width=2.8in]{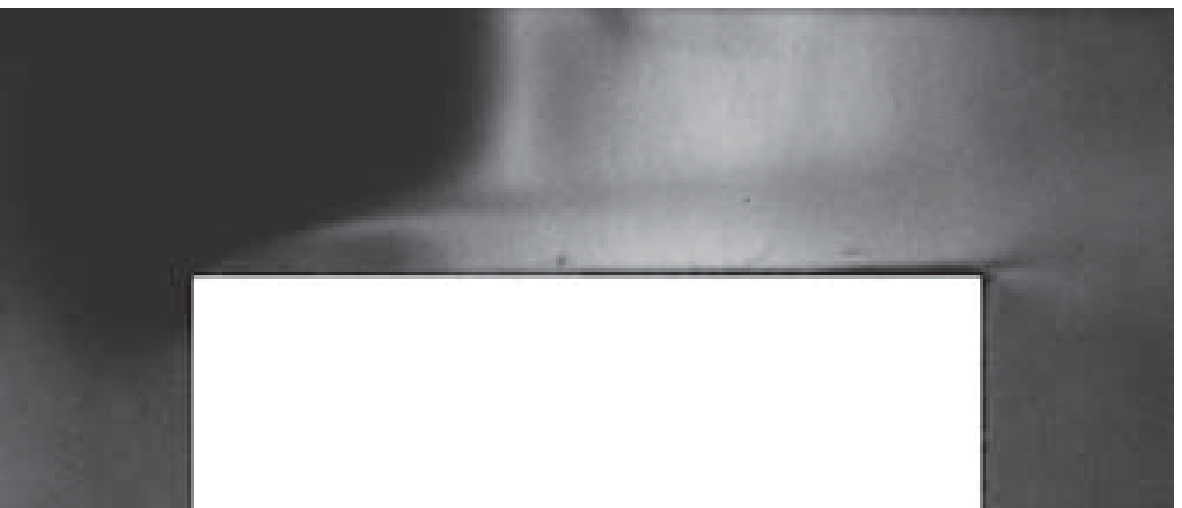} 
		{\\\small(b) VL-cutoff, averaged.}
	\end{minipage}
	\begin{minipage}[t]{0.48\linewidth}
		\centering
		\includegraphics[width=2.8in]{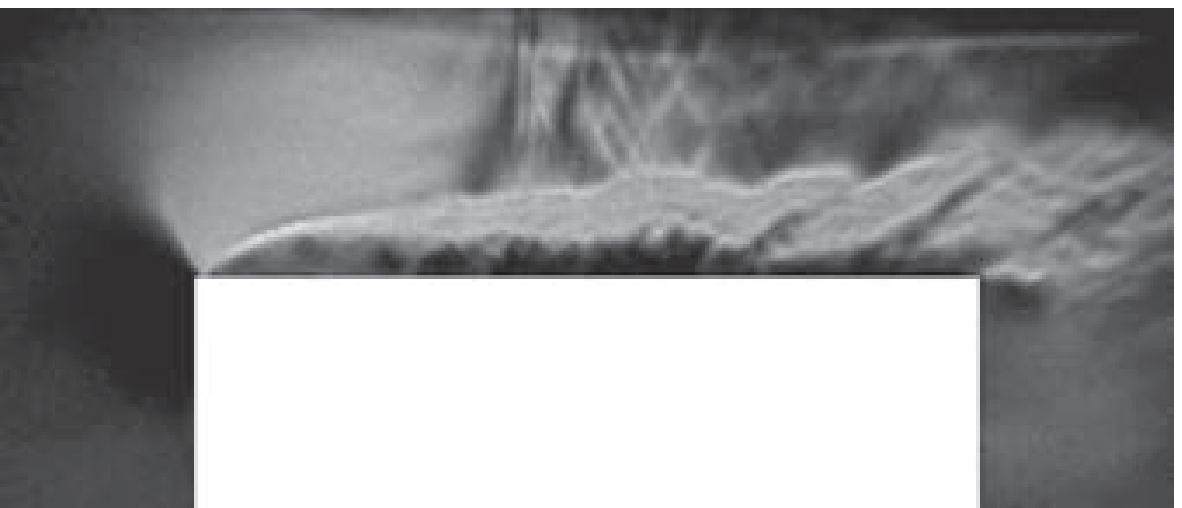}
		{\\\small(c) HD-cutoff, unsteady.}
	\end{minipage}
	\begin{minipage}[t]{0.48\linewidth}
		\centering
		\includegraphics[width=2.8in]{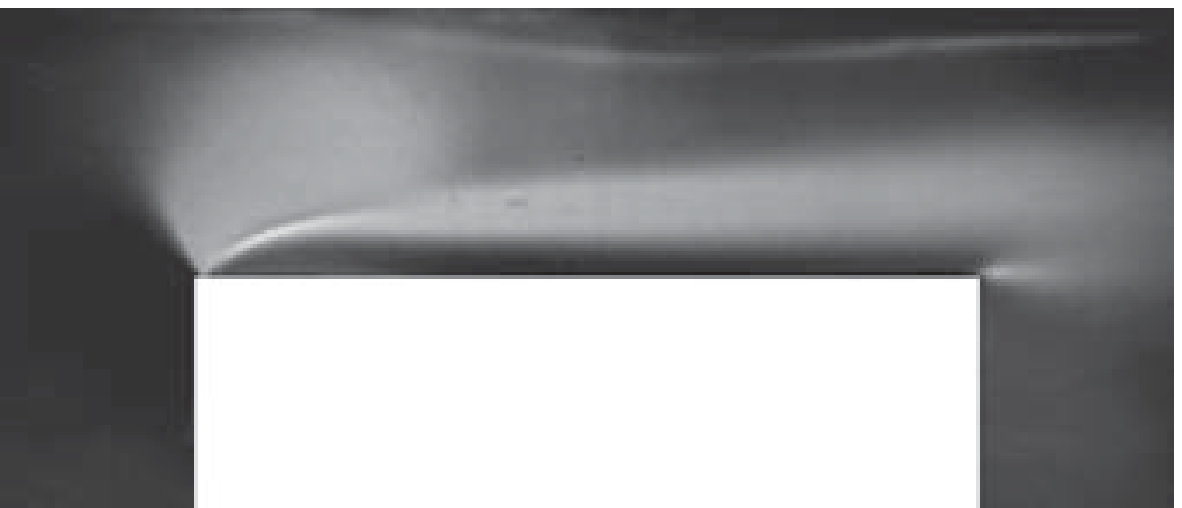}
		{\\\small(d) HD-cutoff, averaged.}
	\end{minipage}
	\caption{\label{g:ref_process3} Compensated unsteady ($g_c$) and averaged ($\bar{g_c}$) grayscale images of the simplified model in the cases of VL- and HD-cutoffs at $r_p=0.297$ (PD120): (a) unsteady view of VL-cutoff \textcolor{red}{(\href{https://pan.baidu.com/s/1werrLJgX3OuOmlGVA9DX3w?pwd=acou}{Multimedia view})}; (b) averaged view of VL-cutoff; (c) unsteady view of HD-cutoff \textcolor{red}{(\href{https://pan.baidu.com/s/1dztyjV75n6F4ndgbFpIVWg?pwd=acou}{Multimedia view})}; (d) averaged view of HD-cutoff. Slope of the model top surface is omitted.}
\end{figure*}
\section{Image postprocessing, pressure distributions, and scale of the separation bubble}\label{s:results1}
\subsection{Postprocessing of schlieren images}
The image postprocessing tasks include averaging, cropping, resizing, and compensating. First, the averaged images are generated from the raw images in grayscale, as calculated by $\bar{g}(i,j)=\sum_{n=1}^{n_p}g_n(i,j)/n_p$, where $g(i,j)$ is the grayscale at pixel index $(i,j)$ and $n_p=$16\,064 is the number of images sampled in a period of $1.888$ s at 8510 images per second. Second, the images are cropped and resized. Third, image compensation is applied to improve the accuracy. The compensated grayscale $g_c$ is calculated from the original grayscale $g$ by
\begin{equation}
g_c=(g-g_b)\frac{\bar{g}_{r}-g_{r,b}}{g_{r}-g_{r,b}},
\end{equation}
where $g_b$, $g_r$, $\bar{g}_r$, $g_{r,b}$ are the grayscale of the background (grayscale measured at the solid model), reference (measured without flow), averaged reference, and reference background, respectively. Examples of the compensated unsteady images are displayed in Figs.~\ref{g:ref_process3}(a) \textcolor{red}{(\href{https://pan.baidu.com/s/1werrLJgX3OuOmlGVA9DX3w?pwd=acou}{Multimedia view})} and \ref{g:ref_process3}(c) \textcolor{red}{(\href{https://pan.baidu.com/s/1dztyjV75n6F4ndgbFpIVWg?pwd=acou}{Multimedia view})}. The averaged images are displayed in Figs.~\ref{g:ref_process3}(b) and \ref{g:ref_process3}(d). 

\begin{figure*}[!htb]
	\begin{minipage}[b]{0.48\linewidth}
		\includegraphics[width=2.8in]{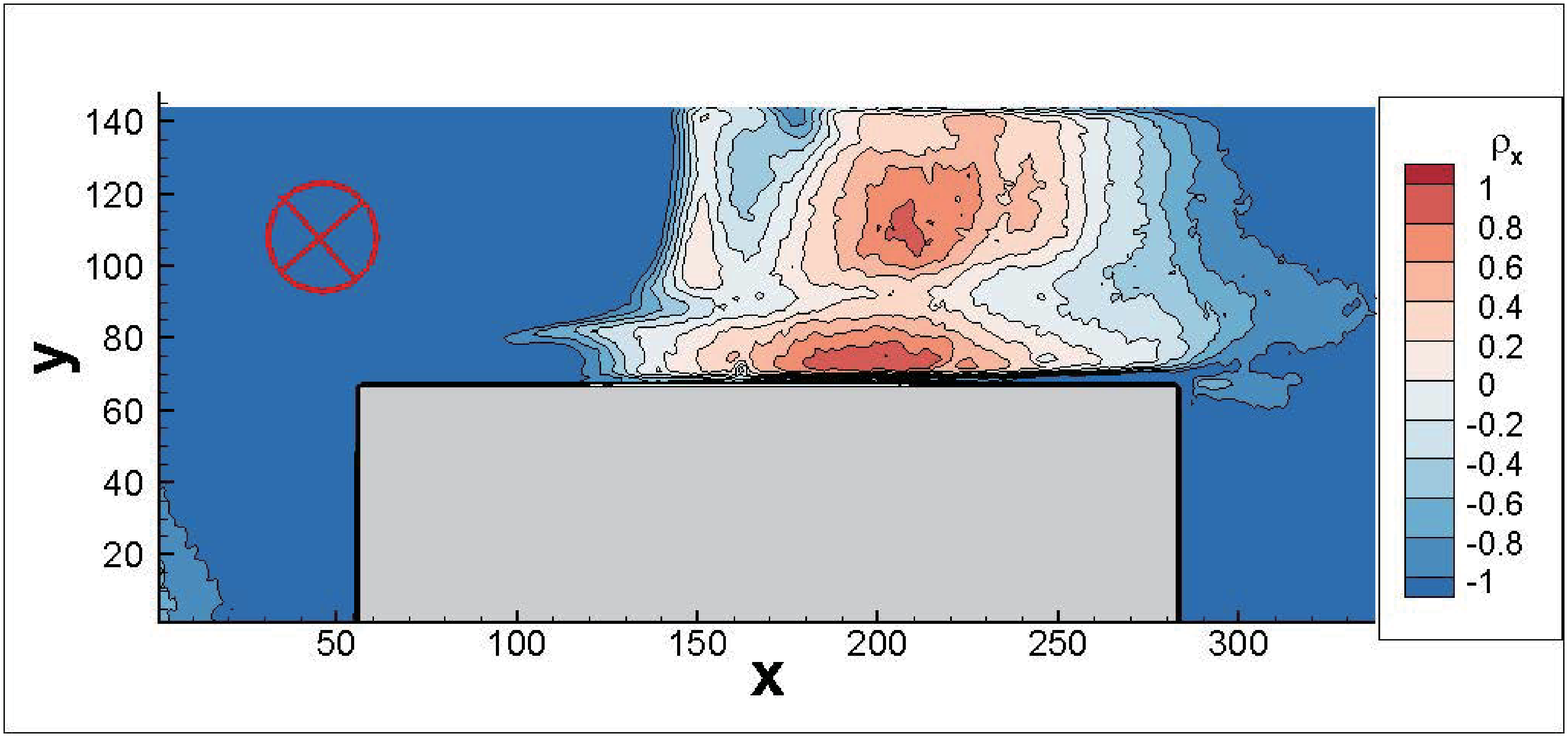}
		{\\(a) VL-cutoff.}
	\end{minipage}
	\hfill
	\begin{minipage}[b]{0.48\linewidth}
		\includegraphics[width=2.8in]{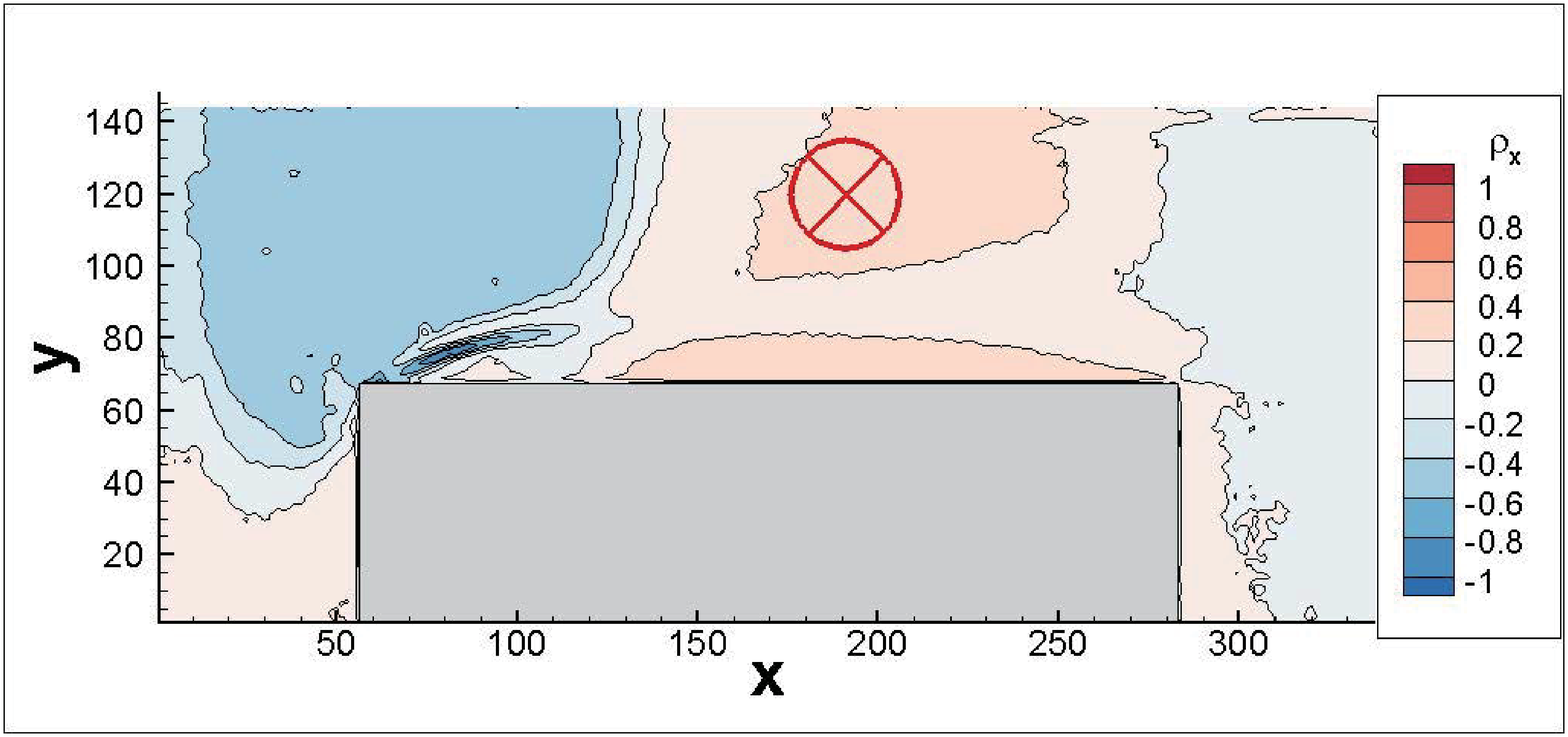}
		{\\(b) VR-cutoff.}
	\end{minipage}
	\vfill
	\begin{minipage}[b]{0.48\linewidth}
		\includegraphics[width=2.8in]{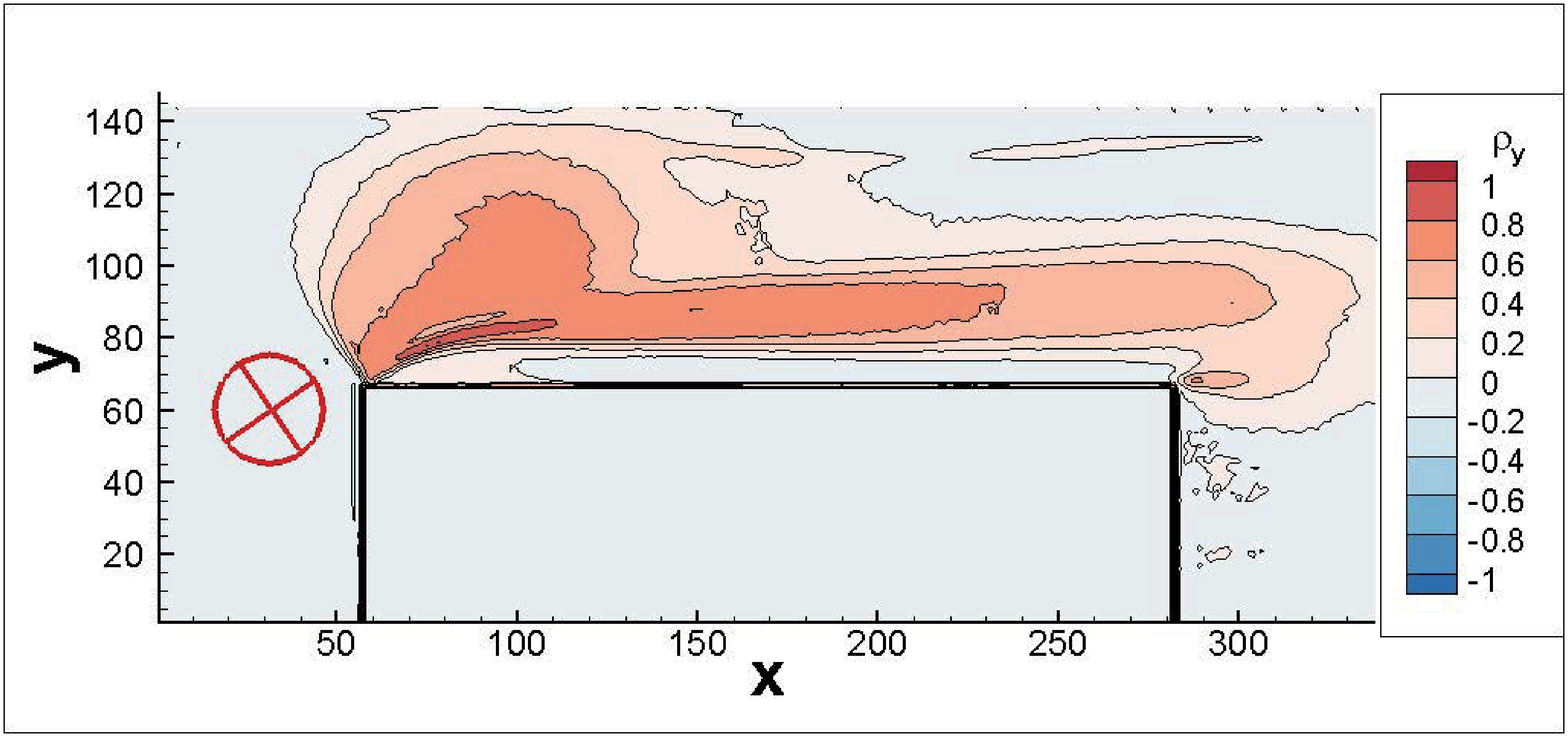}
		{\\(c) HD-cutoff.}
	\end{minipage}
	\hfill
	\begin{minipage}[b]{0.48\linewidth}
		\includegraphics[width=2.8in]{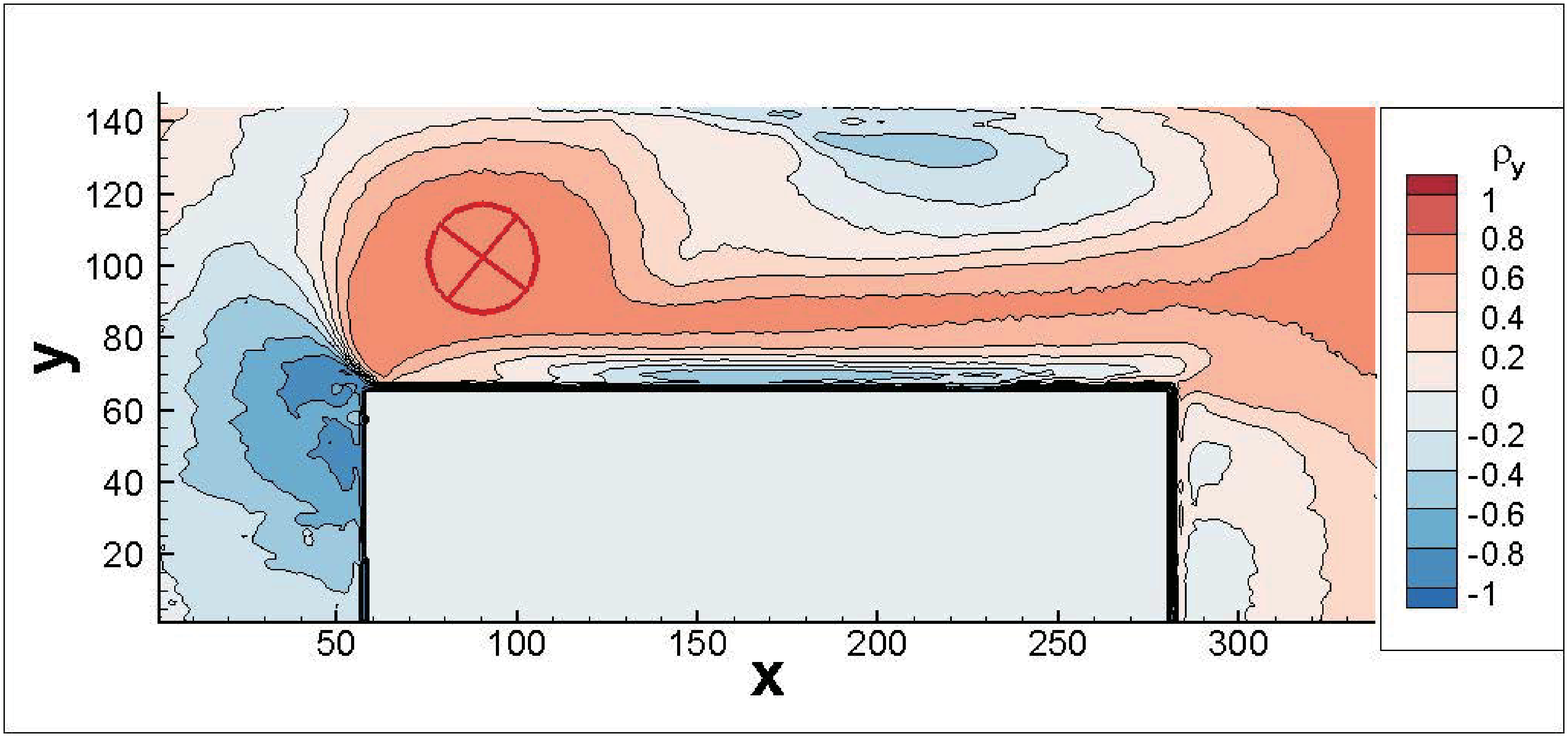}
		{\\(d) HU-cutoff.}
	\end{minipage}
	\caption{\label{g:avg_image_flat120_vl_vr_hd_hu} Averaged density gradient from four cutoff cases at $r_p=0.297$ (PD120) with coordinates in pixels. The invalid region is marked by a circled cross: (a) VL-cutoff; (b) VR-cutoff; (c) HD-cutoff; (d) HU-cutoff.}
\end{figure*}
The averaged pure reference grayscale $\bar{g}_{r,p}=\bar{g}_{r}-g_{r,b}$ is subtracted from the compensated grayscale to obtain a new variable. Due to the linear relationship between the grayscale and the density gradient in the $x$- or $y$-direction, this new variable is termed the scaled density gradient, $K_{sv}\rho^*_x$ or $K_{sh}\rho^*_y$. It is calculated as $K_{sv}\rho^*_x=g_c-\bar{g}_{r,p}~\text{in the case of vertical cutoff}$ and $K_{sh}\rho^*_y=g_c-\bar{g}_{r,p}~\text{in the case of horizontal cutoff}$, where $K_{sv}$ and $K_{sh}$ are the coefficients of the schlieren system in the case of vertical and horizontal cutoffs, respectively. The images are saved with 8-bit unsigned integers (uint8, 0--255); thus, the accuracy is limited to $\pm K_{sv}max(|\rho^*_{x}|)/256$ or $\pm K_{sh}|max(\rho^*_{y})|/256$. Theoretically, if the properties of the light source and the cutoff ratio remain unchanged, the coefficients $K_{sv}$ and $K_{sh}$ should be the same for all cases. However, they are not exactly equal because the dimensions of the light source change slightly when switching from vertical  to horizontal cutoff, and the cutoff ratio is adjusted to give a better view. However, the assertion $K_{sv}\approx CK_{sh}$ is valid, where $C$ is a case-sensitive constant. To eliminate this difference, the density gradients in both directions are normalized based on the maximum value, and the result is recorded as the dimensionless density gradients $\rho_x$ and $\rho_y$.

Limited by the measuring range of the schlieren system, some areas in the sampled images may exceed the measuring range and become invalid, such as the lower-bound areas in the VL- and HD-cutoff cases or the upper-bound areas in the VR- and HU-cutoff cases. Taking the averaged density gradient as an example, as shown in Fig.~\ref{g:avg_image_flat120_vl_vr_hd_hu}, each of these invalid areas is marked by a circled cross. To remove the invalid areas, the overall field of the density gradient is regenerated by the superposition of the VL- and VR-cutoff cases for $\rho_x$, and by the superposition of the HD- and HU-cutoff cases for $\rho_y$. The density gradients in the invalid areas are replaced by the value from the opposite cutoff type. For instance, the invalid area of the VL-cutoff case [see Fig.~\ref{g:avg_image_flat120_vl_vr_hd_hu}(a)] is replaced by that of the VR-cutoff case [see Fig.~\ref{g:avg_image_flat120_vl_vr_hd_hu}(b)]. The average value of these two cases is adopted in the contact area. Thus, the VL- and VR-cutoff cases collapse into the vertical cutoff case, and the HD- and HU-cutoffs turn into the horizontal cutoff case.

\subsection{Pressure distributions and total pressure loss}
\begin{figure}[!htb]
	\centering
	\includegraphics[width=3.in]{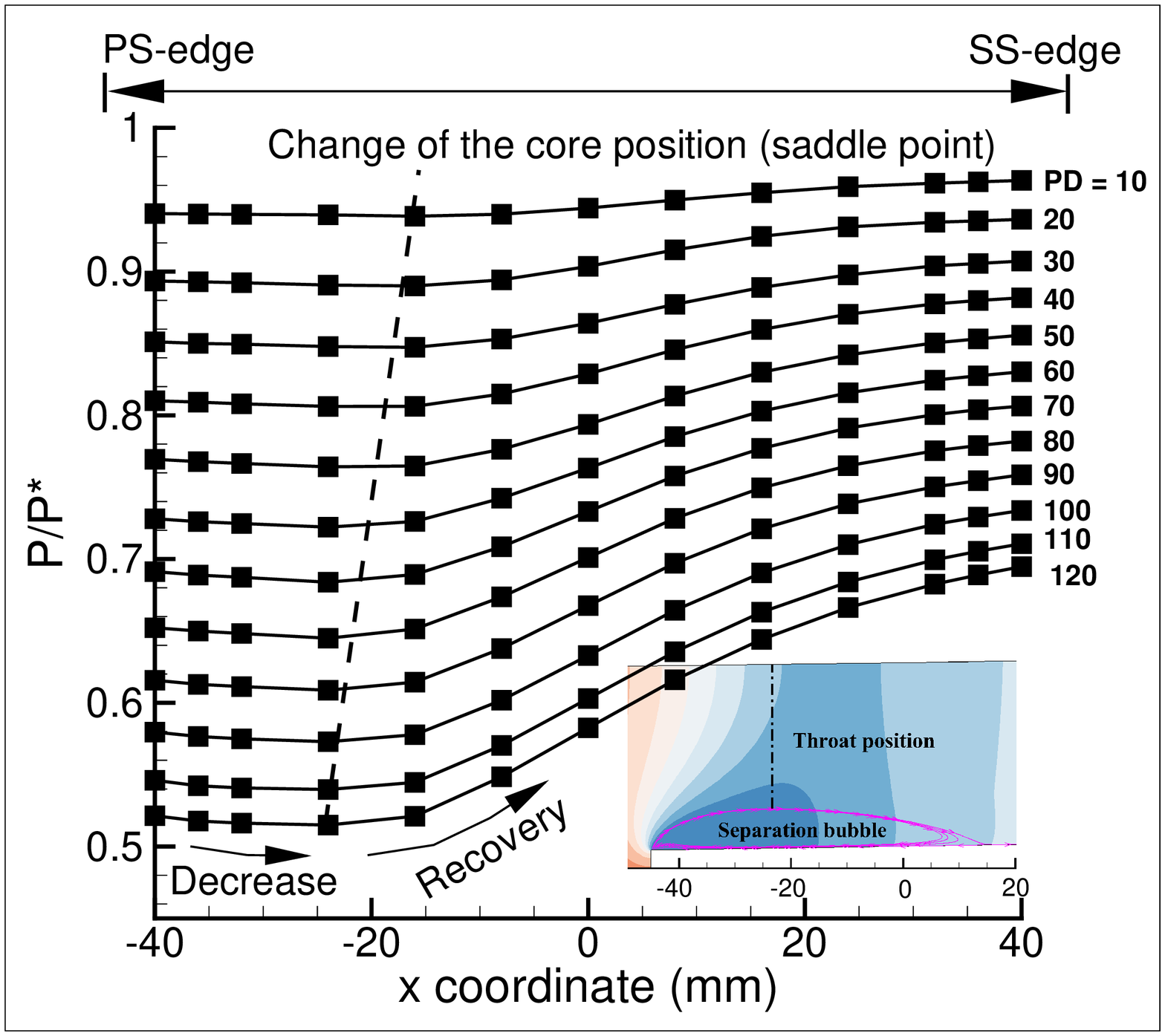}
	\caption{\label{g:sp_vs_x} Pressure distributions along the tip surface under different operating conditions. Curves are labeled by uniformly increasing PD, see details in Table \ref{t:pressure demand}. Corresponding flow ranges from low-subsonic to transonic regime.}
\end{figure}
As shown in Fig.~\ref{g:top_dimension}, five taps are placed along the $Z$-axis to ensure a two-dimensional flow condition. A test run is considered to be valid when the average pressure differences among these taps are less than 5\%. The pressure taps are connected to a Scanivalve pressure scanner that has an accuracy level of $\pm 0.05$\% full-scale or $\pm 2.582$ kPa.

\begin{figure}[!htb]
	\centering
	\includegraphics[width=3.in]{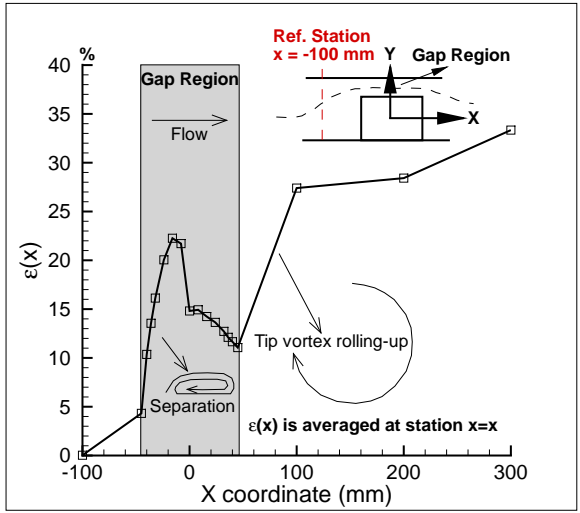}
	\caption{\label{g:total_pressure_loss} Distributions of total pressure loss along the $X$-axis at $r_p=0.297$ (PD120). The total pressure is mass-weighted and averaged at each station $x=x$. Reference station is at $x=-100$ mm, which is $55$ mm upstream of the tip model, and the gap region is between $x=-45$ mm and $x=45$ mm.}
\end{figure}
The averaged static pressures at the taps along the $x$-direction are plotted in Fig.~\ref{g:sp_vs_x}. The pressure is normalized by the inflow total pressure $P^*=101.33$ kPa, and the accuracy level of $P/P^*$ is $\pm 2.55$\%. The static pressure decreases near the PS edge, followed by a region of pressure recovery. The boundary, as indicated by the dashed line, has the lowest pressure, indicating the core position of the separation bubble. Thus, the movement of the saddle point indicates that the separation bubble becomes smaller as the blade loading increases.

The mass-weighted-average total pressure at section $x=x$ is termed $\bar{P}^*(x)$. Thus, the section-averaged total pressure loss ratio is defined as
\begin{equation}
\label{e:tp_loss}
\epsilon(x) = \frac{\bar{P}_{ref}^*-\bar{P}^*(x)}{\bar{P}_{ref}^*}\times 100\%,
\end{equation}
where $\bar{P}_{ref}^*$ is the reference total pressure averaged at $x=-100$ mm. Numerical simulations were performed to evaluate the total pressure loss at $r_p=0.297$ (PD120). As shown in Fig.~\ref{g:total_pressure_loss}, the total pressure loss increases along the $X$-direction. Most of this loss occurs in the separation and the tip-vortex rolling-up regions. The accumulated loss in the separation zone is about 18\%, and the zone near the tip-vortex rolling-up region has a similar loss range. In addition, the total pressure recovers by approximately 12\% near the exit of the tip clearance. Nearly 30\% of the total pressure loss occurs around the gap region.

\subsection{Scale of the PS separation bubble }
\begin{figure}[!htb]
	\centering
	\includegraphics[width=3.in]{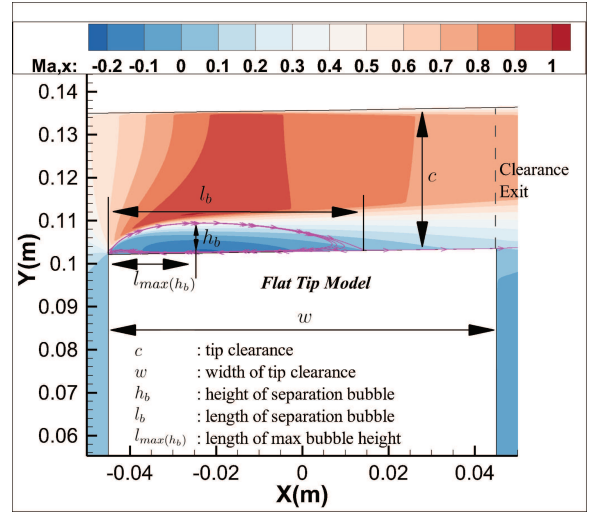}
	\caption{\label{g:separation_bubble_size} Definition of the size and location of the separation bubble defined by the PS recirculating zone, flooded by the averaged Mach number in the $x$-direction at $r_p=0.297$ (PD120). The separation bubble becomes reattached, which is consistent with Storer's observations \cite{storer1991} for $c/w\le40\%$.}
\end{figure}

\begin{figure}[!htb]
	\centering
	\includegraphics[width=3.in]{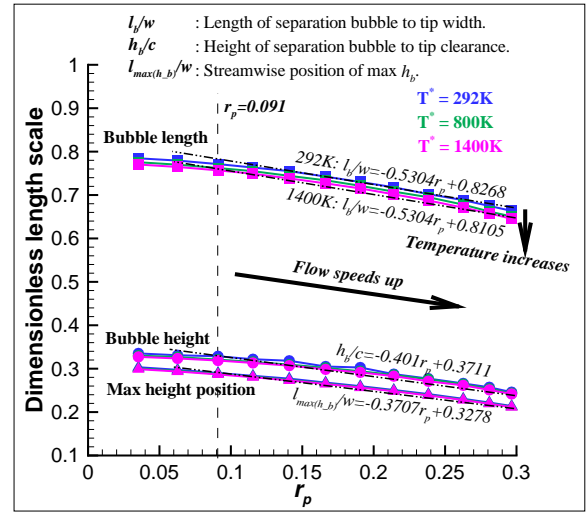}
	\caption{\label{g:separation_bubble_length_scale} Dimensionless size and location of separation bubble. [The contraction ratio is theoretically given by $\sigma=\pi/(\pi+1)\approx 0.611$ for a potential flow \cite{milne-thompson1968}. The contraction ratio measured herein ($\sigma=1-h_b/c \in [0.665,0.754]$) is larger due to flow compressibility.]}
\end{figure}

The PS separation bubble is a critical feature of TLFs. The size of the separation bubble is defined in Fig.~\ref{g:separation_bubble_size}. The length scales are made dimensionless according to the width or height of the tip clearance. As shown in Fig.~\ref{g:separation_bubble_length_scale}, the size of the separation bubble decreases almost linearly with increasing blade loading. This explains the movements of the pressure saddle point shown in Fig.~\ref{g:sp_vs_x}. Temperature has some weak effects on the bubble length. 
A higher operating temperature leads to a shorter separation bubble, although the slope with respect to the pressure ratio is unchanged. Moreover, the height and position of maximum height remain almost constant under different operating temperatures. This feature is important for turbine flows in which it is important to control the blade heat load.

\begin{figure*}[!htb]
	\begin{minipage}[t]{0.48\linewidth}
		\centering
		\includegraphics[width=2.8in]{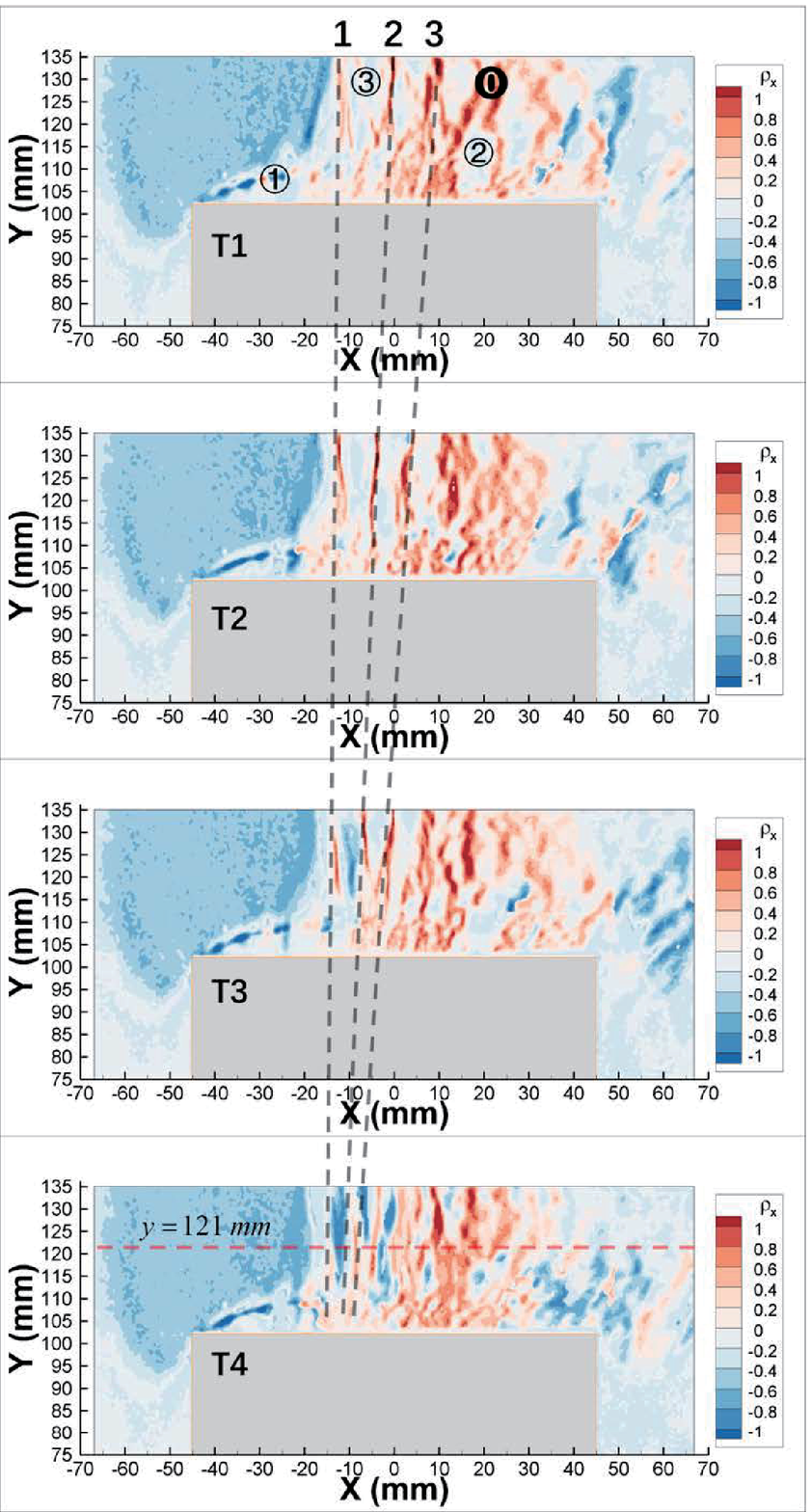} 
		{\\\small(a) Measured $\rho_x$ by vertical cutoff.}
	\end{minipage}
	\begin{minipage}[t]{0.48\linewidth}
		\centering
		\includegraphics[width=2.8in]{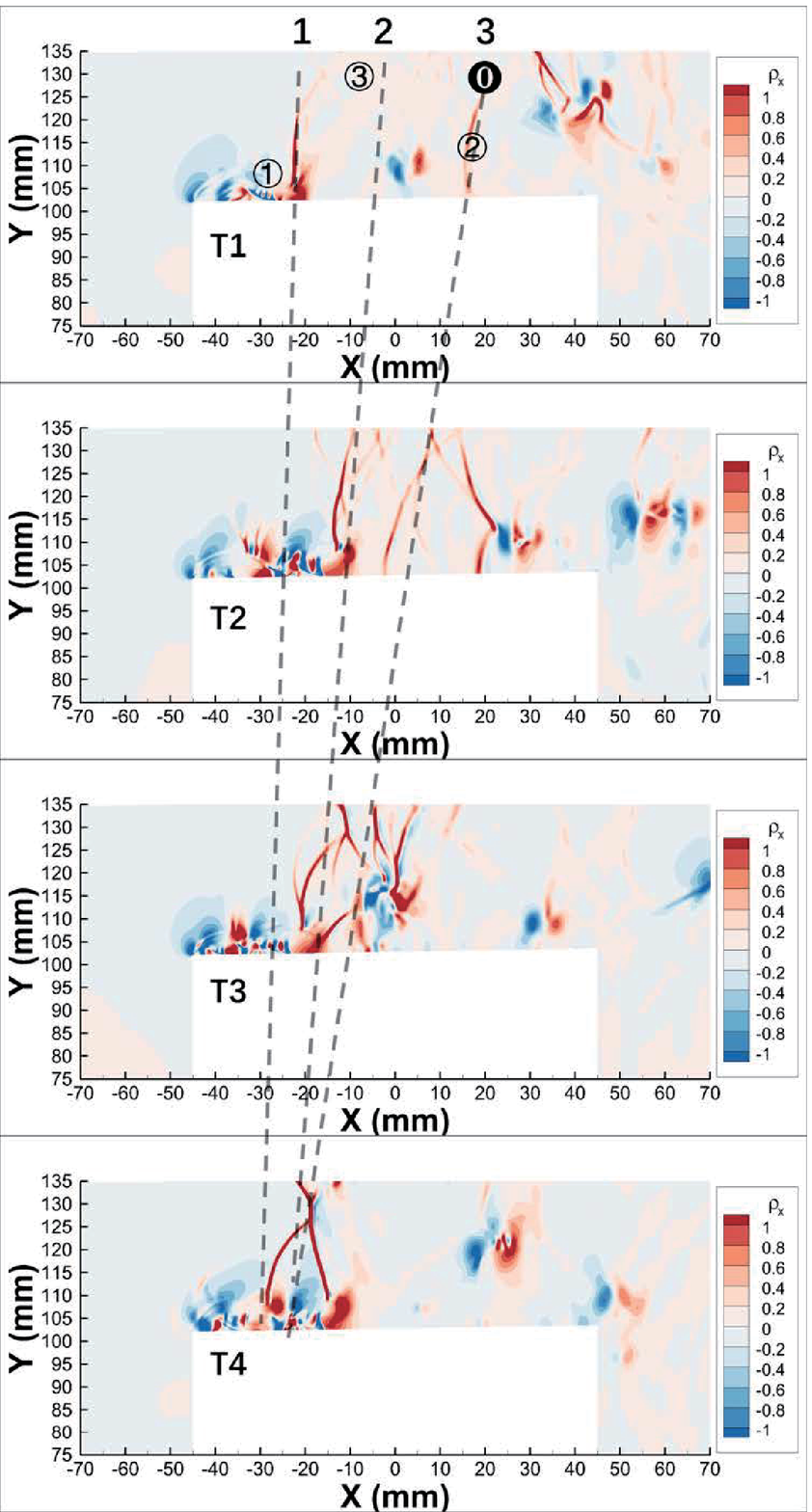}
		{ \\\small(b) Predicted $\rho_x$ by IDDES.}
	\end{minipage}
	\caption{\label{g:history_flat120_vl_vr} Time-resolved density gradient distributions at $r_p=0.297$ (PD120): (a) measured $\rho_x$; (b) predicted $\rho_x$. Shock motions are indicated by dashed lines. The starting time is T1$\approx5$~s in the experiment and T1$=0.014165$~s for the simulation. The time interval in both cases is $\delta t=1/8510$~s$\approx1.175\times 10^{-4}$~s. Two-dimensional effects lead to a stronger shock with a higher moving velocity (smaller dashed-line slope).}
\end{figure*}

\section{Verification of schlieren-observed flow structures and effects of pressure difference}\label{s:results2}
First, the results from the extreme operating condition ($r_p=0.297$) are analyzed. The observed flow structures and their motions are cross-verified by numerical simulations. Second, the effects of pressure differences are discussed.

\begin{figure}[!htb]
	\centering
	\includegraphics[width=3.4in]{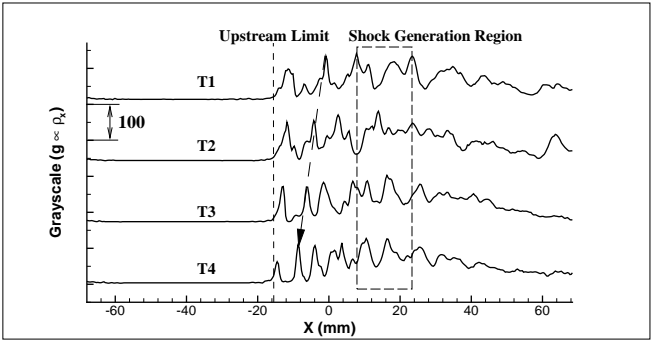}
	\caption{\label{g:ustd_plot_flat120vl}Measured grayscale along line $y=121$ mm [see Fig.~\ref{g:history_flat120_vl_vr}(a)-T4] with VL-cutoff at $r_p=0.297$ (PD120). Curves are evenly redistributed in the vertical direction with an increment of 170. Timesteps T1--T4 are the same as in Fig.~\ref{g:history_flat120_vl_vr}.}
\end{figure}

Figure \ref{g:history_flat120_vl_vr}(a) displays four time steps of the unsteady flow measured with a vertical cutoff at $r_p=0.297$ (PD120). In the region labeled \circled{1}, several vertical-line structures are present, and their motions are marked by dashed lines. These structures are verified to be shock waves using the IDDES simulations, as shown in Fig.~\ref{g:history_flat120_vl_vr}(b). Unstable shock waves are generated in the compression region behind the third shock, labeled \circled{3} at timestep T1. As the shock wave grows stronger, it propagates upstream with decreasing strength and travel speed. This process continues until the shock wave stops moving and gradually fades out when reaching the upstream limit. A separation bubble labeled \circled{2} is formed near the pressure-side edge. The low-high-couple pattern at the boundary of the separation bubble indicates the vortex shedding that generates the main vortex street. A shear layer wraps around the separation bubble and is convected downstream. The periodic vortex shedding and the resulting shear layer flapping are the main driving forces in the unsteadiness of the shock waves. When the shock waves reach the upstream limit, as marked by the first dashed line, they become relatively steady due to the reduced unsteadiness at the front of the separation bubble. Thus, the upstream limit of the over-tip shock waves is dominated by the mean flow. The generation and evolution of over-tip shock waves is more clearly captured by the footprints along the line $y=121$ mm, as plotted in Fig.~\ref{g:ustd_plot_flat120vl}. Each of the first few spikes indicates a shock wave. These shock waves are generated in the region enclosed by the rectangular box. After their formation, the shock waves travel upstream and gradually fade out until they vanish at the upstream limit.

\begin{figure*}[!htb]
	\begin{minipage}[t]{0.48\linewidth}
		\centering
		\includegraphics[width=2.8in]{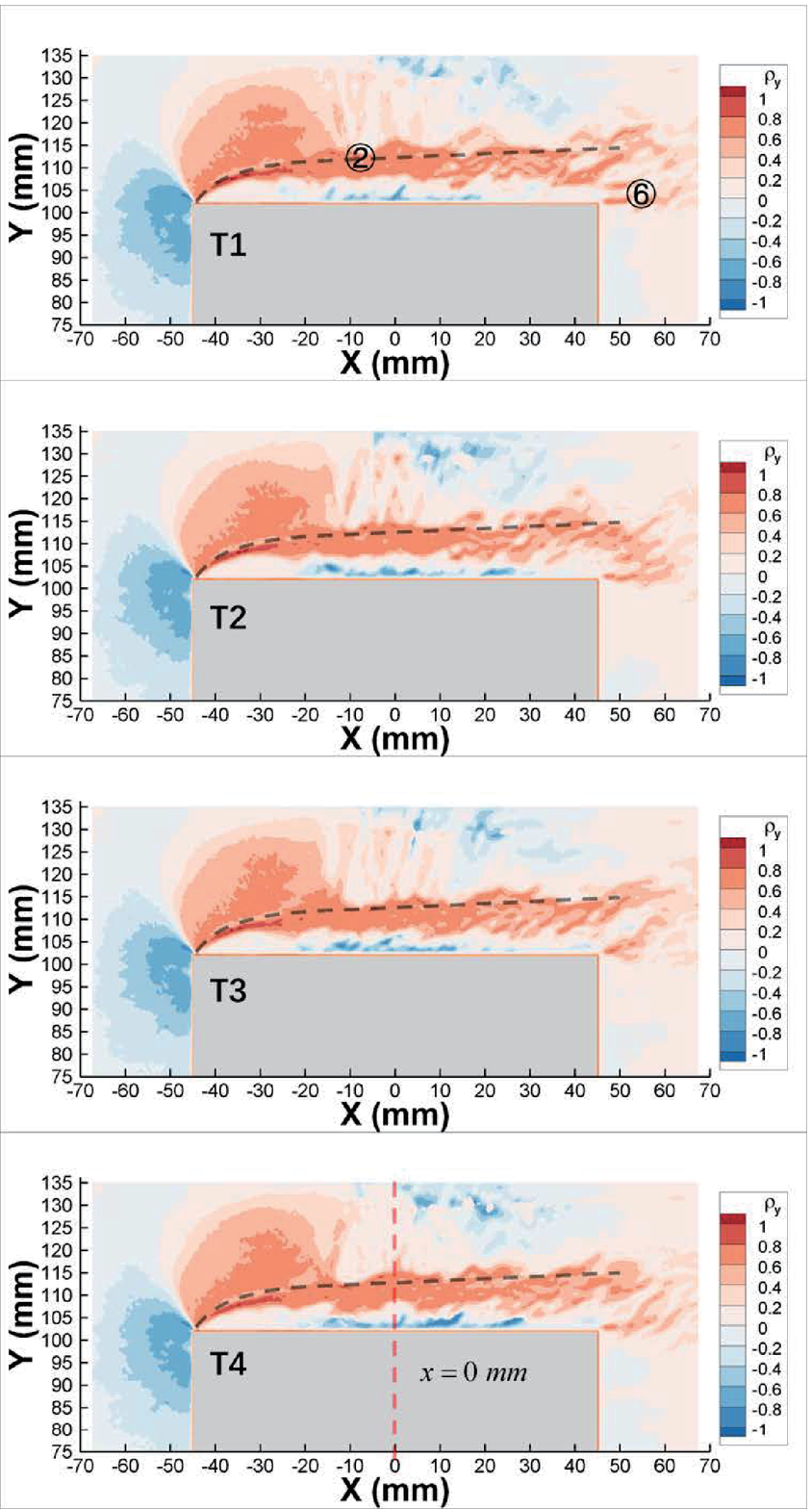} 
		{\\\small(a) Measured $\rho_y$ by horizontal cutoff.}
	\end{minipage}
	\begin{minipage}[t]{0.48\linewidth}
		\centering
		\includegraphics[width=2.8in]{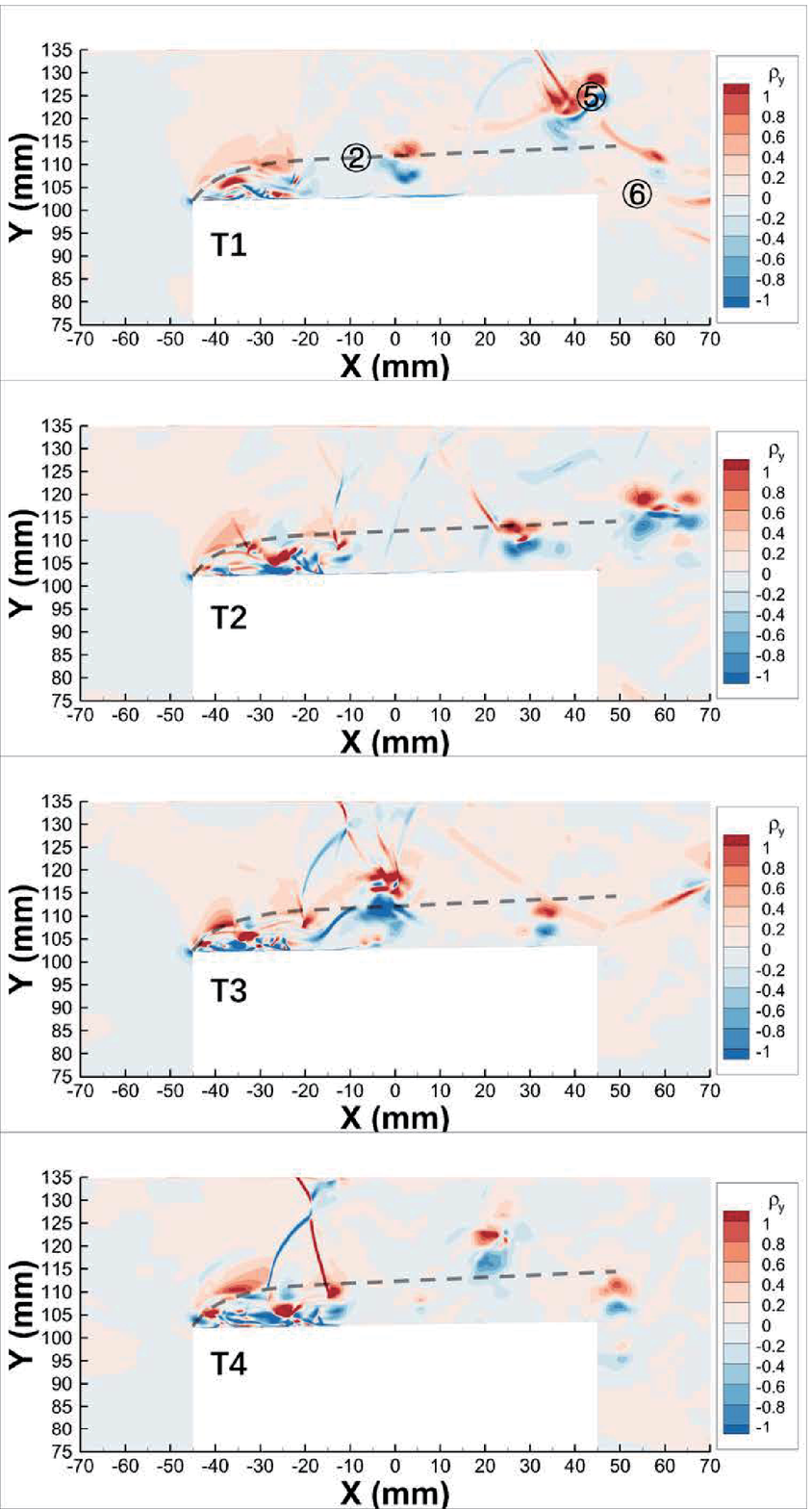}
		{ \\\small(b) Predicted $\rho_y$ by IDDES.}
	\end{minipage}
	\caption{\label{g:history_flat120_hd_hu} Time-resolved density gradient distributions at $r_p=0.297$ (PD120): (a) measured $\rho_y$; (b) predicted $\rho_y$. The dashed curves in black are the shear lines measured in Fig.~\ref{g:separation_bubble_size}. The starting time is T1$\approx5$~s in the experiment and T1$=0.014165$~s for the simulation. The time interval in both cases is $\delta t=1/8510$~s$\approx1.175\times 10^{-4}$~s.}
\end{figure*}
\begin{figure}[!htb]
\centering
\includegraphics[width=3.4in]{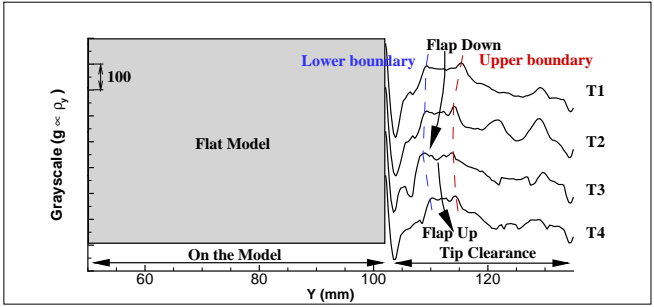}
\caption{\label{g:ustd_plot_flat120hd}Measured grayscale, $g\propto \rho_y$, along the line $x=0$ mm [see Fig.~\ref{g:history_flat120_hd_hu}(a)-T4] with VL-cutoff at $r_p=0.297$ (PD120). To illustrate the shear flapping over time, the curves are evenly redistributed in the vertical direction with an increment of 170. T1--T4 are the same as in Fig.~\ref{g:history_flat120_hd_hu}.}
\end{figure}
The horizontal flow structures are captured by the horizontal schlieren-cutoff. As shown in Fig.~\ref{g:history_flat120_hd_hu}, flapping motions of the flow structures labeled \circled{2} and \circled{5} can be observed. These structures are verified by the IDDES results to be vortex shedding and the over-tip shear layer. The upper boundary of the observed flapping flow is consistent with the shear layer in Fig.~\ref{g:separation_bubble_size}(a). Thus, flow structure \circled{2} is the shear layer, and it contains a vortex street. Additionally, a small shear layer labeled \circled{6} is generated from the suction-side edge. This is the vortex shedding induced by the Tollmien--Schlichting instability. The flapping of the shear layer is clearly revealed by the footprints along the line $x=0$ mm plotted in Fig.~\ref{g:ustd_plot_flat120hd}. The high platforms indicate the region of the vortex street and the formation of the shear layer. The upper boundary coincides with the shear layer of $V_x$, and the center line of these high platforms is composed of the loci of vortex cores. The shear layer and vortex street are observed to flap up and down.

\begin{figure*}[!htb]
\begin{minipage}[t]{0.48\linewidth}
\centering
\includegraphics[width=2.8in]{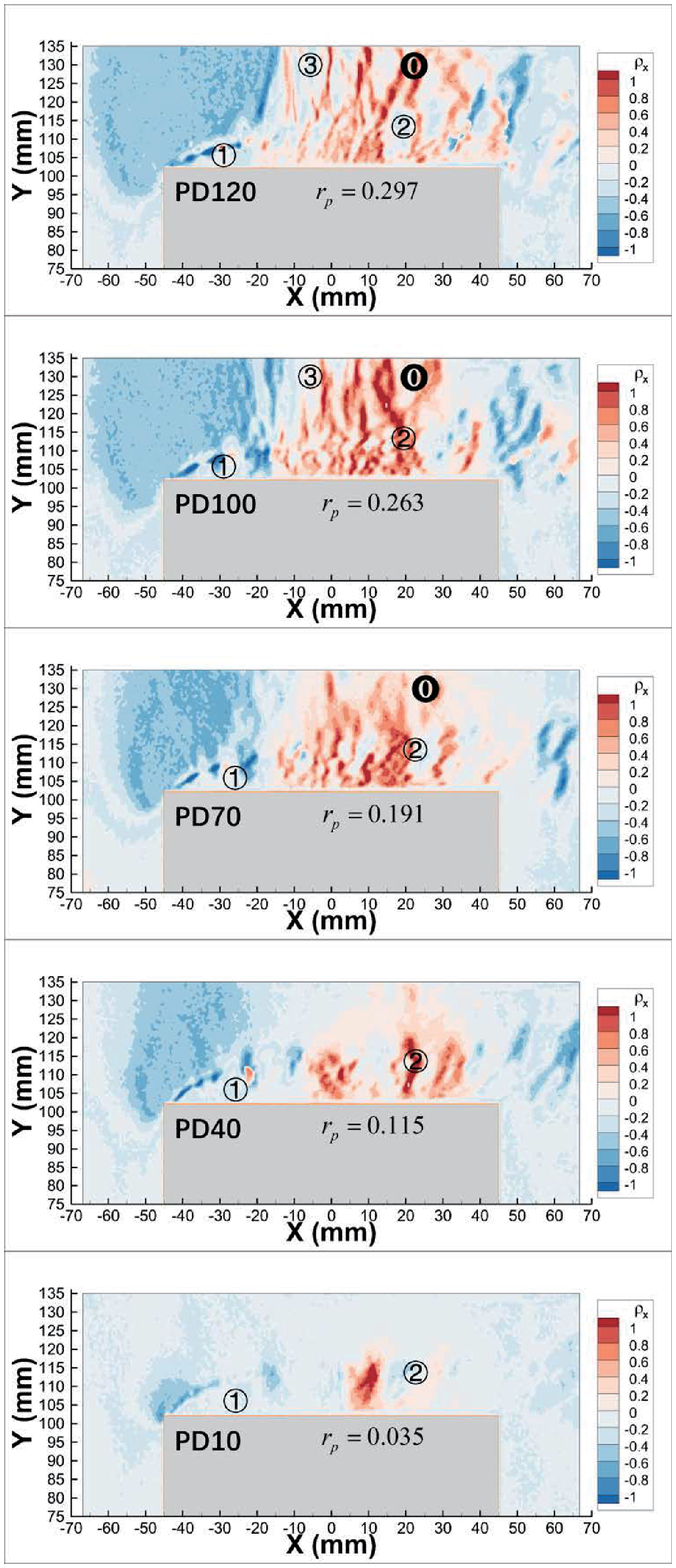}\\
{\small {(a) Vertical cutoff.}}
\end{minipage}
\begin{minipage}[t]{0.48\linewidth}
\centering
\includegraphics[width=2.8in]{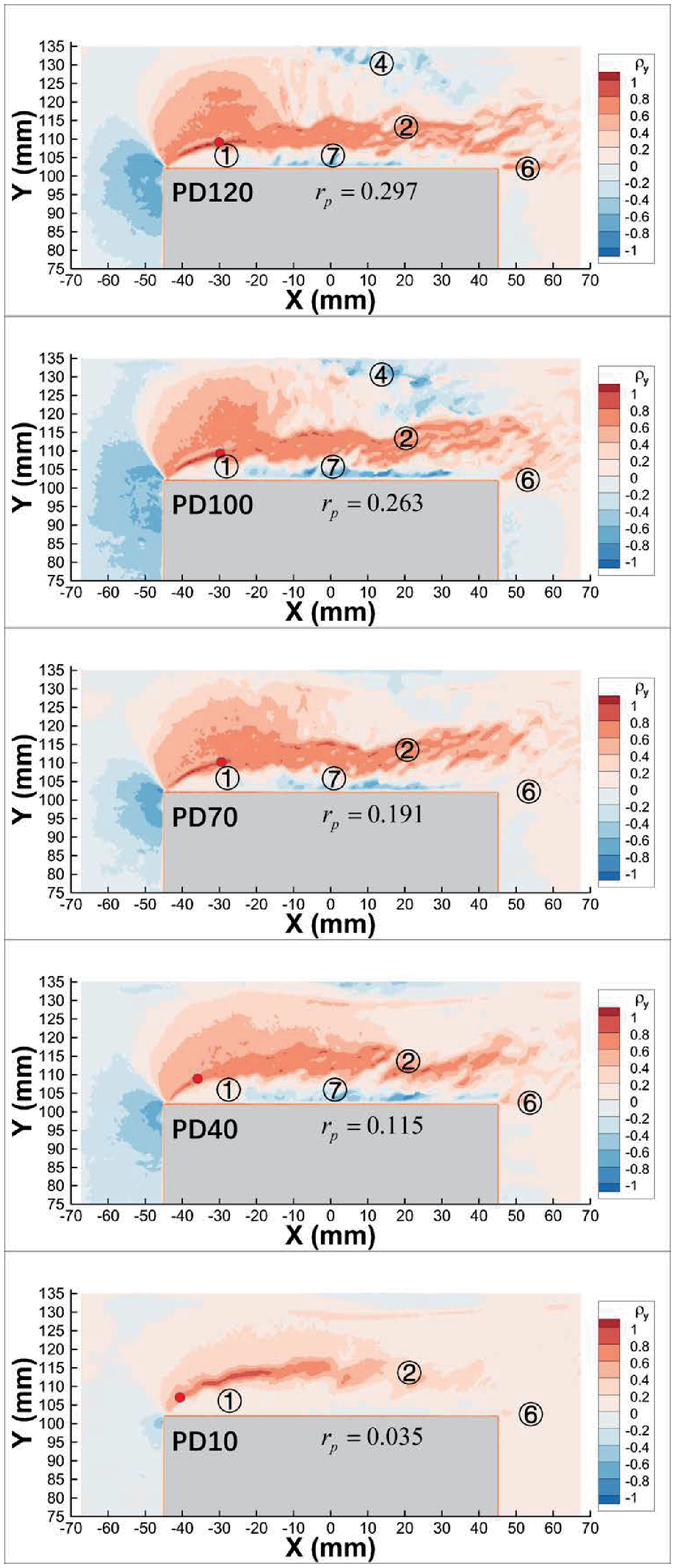}\\{\small {(b) Horizontal cutoff.}}
\end{minipage}
\caption{\label{g:history_flat_pds} Snapshots of $(\rho_x,\rho_y)$ at five operating conditions: (a) $\rho_x$ measured by vertical cutoff; (b) $\rho_y$ measured by horizontal cutoff. Values are normalized based on case PD120 ($r_p=0.297$). Flow structures are indicated by circled numbers: \circled{1} PS-edge separation bubble; \circled{2} main vortex street and shear layer; \circled{3} oscillating shock waves; \circled{4} shock-induced vortices; \circled{5} escaped PS-edge vortices from the vortex street (reserved); \circled{6} SS-edge vortex shedding; \circled{7} secondary vortices induced by the vortex street.}
\end{figure*}
The tip clearance flow structures rely on the pressure difference between the pressure and suction sides. Higher pressure differences indicate higher blade loading and a larger pressure ratio $r_p$. Figure \ref{g:history_flat_pds} shows the effects of the pressure difference on TLF structures, which are labeled by circled numbers following the verification in Figs.~\ref{g:history_flat120_vl_vr} and \ref{g:history_flat120_hd_hu}. Low-subsonic ($r_p \in [0.0,0.063]$), subsonic [$r_p\in (0.063,0.263)$], and transonic ($r_p\in [0.263,0.297]$) operating conditions are covered. As shown in Fig.~\ref{g:history_flat_pds}(a), only two flow structures are revealed by PD10 when the flow compression is weak. As the blade loading increases, compressibility gradually takes control of the flow in the region near the casing wall. Shock waves start to form in region \circled{0} at PD70 ($r_p=0.191$). These are not consistently visible before PD100 ($r_p=0.263$), when the propagation of shock waves can be observed at \circled{3}. Going beyond PD100, the over-tip shock waves grow stronger and thinner, and can travel further upstream. At the same time, there are interactions between the compression waves and the shear layer, resulting in the suppression of the separation bubble (labeled \circled{1}) and the breakdown of the vortex street. The shear layers and related flow structures are highlighted by the vertical cutoff, as shown in Fig.~\ref{g:history_flat_pds}(b). Three flow structures are revealed by PD10, and sufficiently strong secondary and shock-induced vortices appear as the pressure difference increases. The shear layer is initially stable before the red dot, as shown in Fig.~\ref{g:history_flat_pds}(b). This red dot in each subfigure denotes the starting point of the shear-layer unsteadiness. The dots move downstream under subsonic conditions as the pressure ratio increases; however, this pattern is reversed when shock waves emerge after $r_p=0.263$ (PD100). This indicates that flow acceleration helps stabilize the foremost part of the shear layer and the separation bubble. One possible reason for this phenomenon is the ability of over-tip acceleration to suppress the upstream turbulence \cite{wheeler2016}. This mechanism is disturbed by the motion of over-tip shock waves, resulting in a smaller stabilized region under transonic conditions. Thus, the unsteady feature of over-tip shock waves is a key factor controlling the instability of TLFs under transonic conditions.

\section{Lock-in effect of over-tip shock waves and identification of the escaping vortex-shedding mode}\label{s:results3}
The vortex-shedding and shear-layer flapping motions, as illustrated by Fig.~\ref{g:history_flat_pds}, are primary flow patterns in tip clearance flows that exist under low-subsonic, transonic, and supersonic conditions. \cite{wheeler2016} (The index \circled{2} is assigned to these flow structures.) When the TLF is accelerated to the transonic regime, it becomes highly unstable, resulting in the generation and evolution of over-tip shock waves, the generation of shock-induced boundary separation, and a new escaping vortex-shedding flow. Fourier analysis and DMD can be used to understand such flow features. The transonic operating condition, i.e., PD120 with $r_p=0.297$, covers most of the aforementioned flow structures and their changes with respect to blade loading. Thus, we consider transonic conditions as a representative of TLFs.
\subsection{Fourier and DMD modes}

\begin{figure}[!htb]
\centering
\includegraphics[width=3.in]{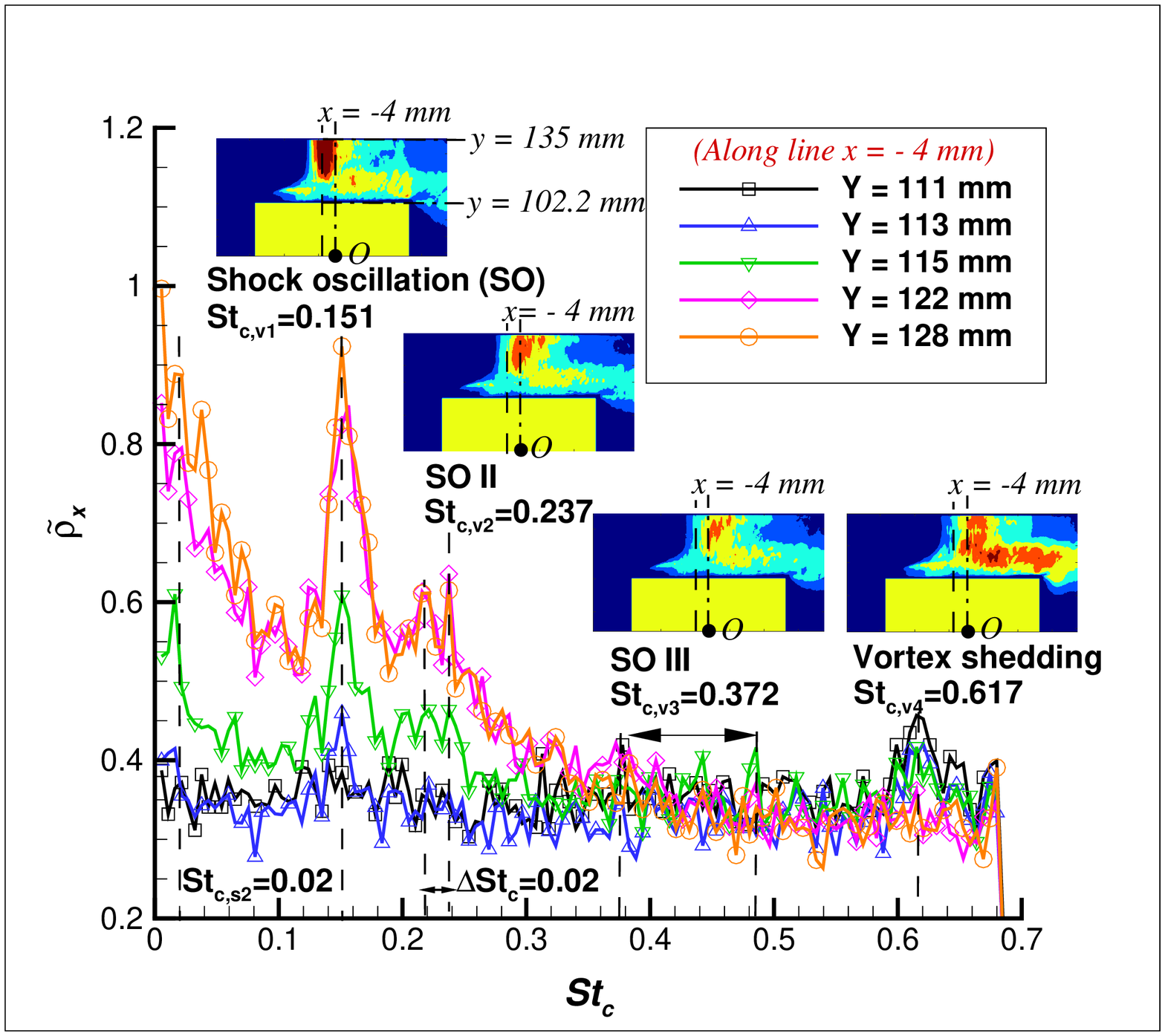}
\caption{\label{g:spectra_along_line_flat_vl} Spectra and dominant Fourier mode patterns (embedded subfigures) of $\rho_x$ measured by vertical cutoff at $r_p=0.297$ (PD120). Amplitude is normalized by the maximum peak. Vertical line $x=-4$ mm is the location of the strongest shock oscillation. Peaks at $St_{c,s1}=0.01$ and $St_{c,s2}=0.02$ are system vibrations.}
\end{figure}
\begin{figure}[!htb]
\centering
\includegraphics[width=3.in]{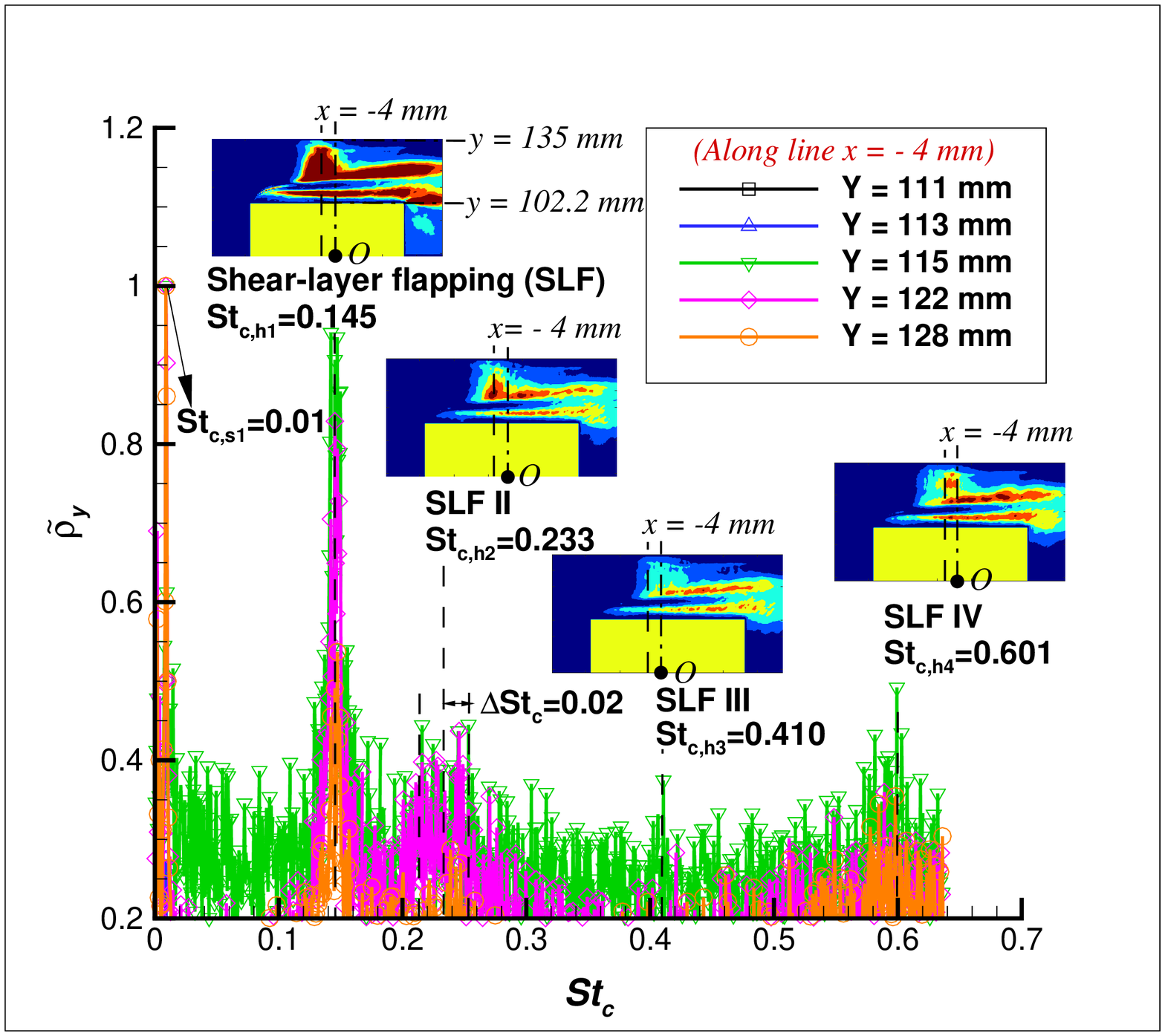}
\caption{\label{g:spectra_along_line_flat_hd} Spectra and dominant Fourier mode patterns of $\rho_y$ measured by horizontal cutoff at $r_p=0.297$.}
\end{figure}

First, Fourier transforms are applied pixel-by-pixel to the measured images. The locations and frequencies of the periodic flow structures are revealed by the resulting Fourier modes. As shown in Fig.~\ref{g:spectra_along_line_flat_vl}, three shock oscillation (SO) modes are present on the upper part of the line $x=-4$ mm. The Strouhal number is defined as $St_c=fc/V_{x,avg}$, with $V_{x,avg}\approx206.7$ m/s being the averaged $x$-velocity at the gap exit and $c=32.8$ mm being the clearance. The strength of the SO rises as the $y$-coordinate increases. In the lower part of the line $x=-4$ mm, a vortex shedding mode appears when $y\le 115$ mm. Distinct SO and vortex-shedding modes are visible at $(x,y)=(-4,115)$ mm. This indicates that shock/shear-layer/vortex-shedding interactions occur in this region. The three observed SO modes are consistent with the first three shear-layer flapping (SLF) modes in Fig.~\ref{g:spectra_along_line_flat_hd}, and the vortex-shedding mode coincides the fourth SLF mode. This implies that the SO and vortex-shedding modes are almost locked-in with the SLF modes through these interactions. Therefore, only four independent modes are observed. Nevertheless, the second SO and SLF modes are unstable. The intermittency of these structures leads to side-lobe modes with $St_{c,v3}\pm \Delta St_c$. The reason is the interaction with the second harmonic of the system vibration at $St_{c,s2}=0.02$. A different intermittency pattern is observed at the third SO mode; this is discussed later and identified as the escaping vortex mode. 

\begin{figure}[!htb]
\centering
\includegraphics[width=3.in]{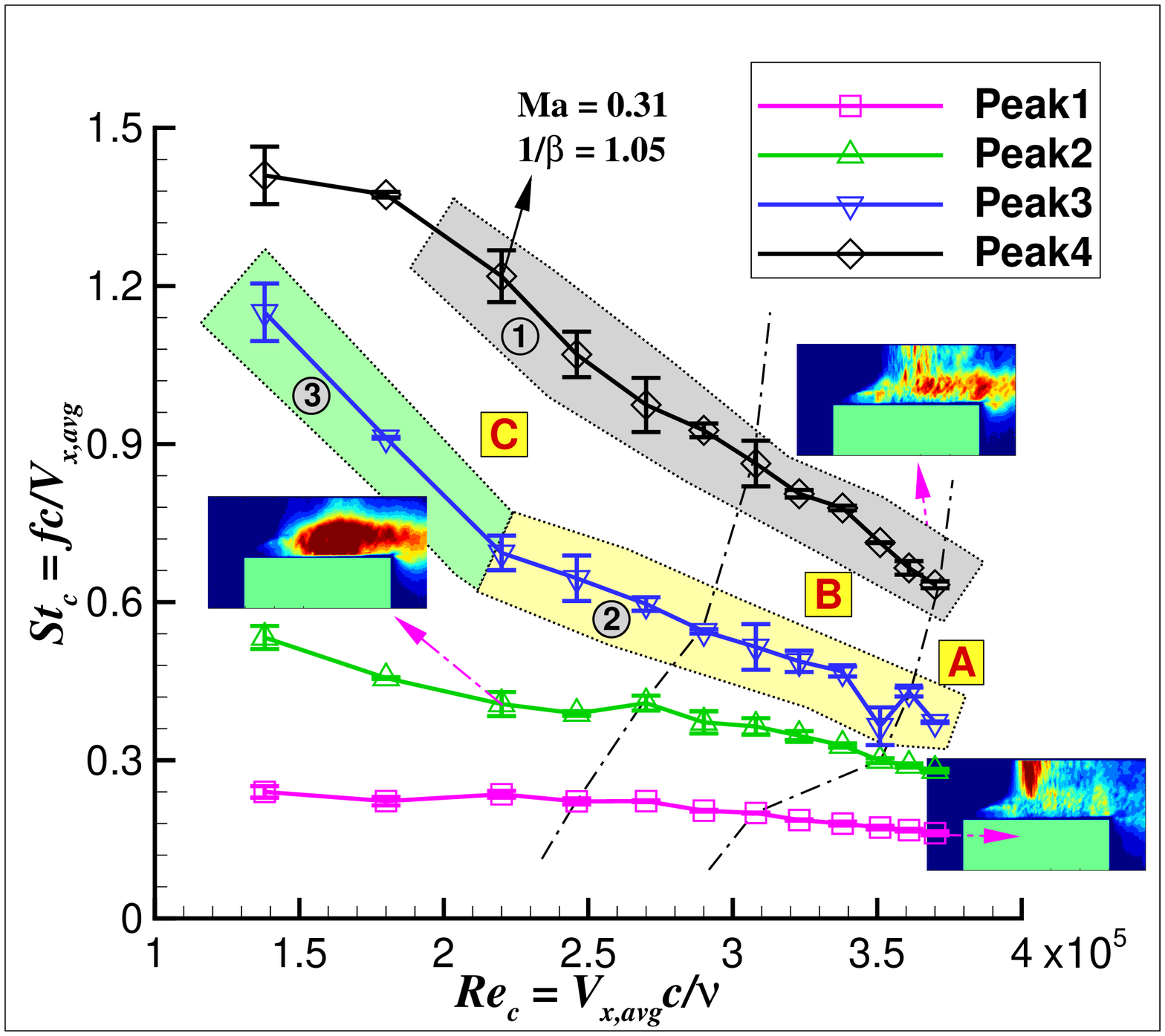}
\caption{\label{g:eigen_frequencies_vs_pds} Change of four Fourier modes with Reynolds number. Squared letters indicate three zones split by dash-dotted lines. Zone A: dominated by shock/compression waves. Zone B: dominated by vortex--shock coupling. Zone C: dominated by vortex shedding. Circled numbers indicate curve segments that carry the same flow mechanism. Error bars are based on the frequency width of the narrowband peaks.}
\end{figure}
Four peak frequencies can be observed under subsonic conditions when the first three modes are dominated by vortex shedding instead of shock oscillation. As shown in Fig.~\ref{g:eigen_frequencies_vs_pds}, the peak frequencies change with the Reynolds number. As indicated by the dash-dotted lines, these curves fall into three zones, A, B, and C. The SO mode is observed in zone A. The interaction of SO with vortex-shedding modes is observed in zone B. An almost-pure vortex-shedding mode is observed in zone C. Near-constant patterns (flat lines) of peak 1 and peak 2 are observed, which implies that steady coupling of vortex shedding and the oscillation of shock/compression waves is established. As indicated by curve segments \circled{2} and \circled{3}, the flow mechanism of peak 3 switches when the pressure ratio increases beyond $r_p=0.063$. This switching is another symptom of the intermittency feature, as observed by the third peak in Fig.~\ref{g:spectra_along_line_flat_vl}. Additionally, the frequencies of peak 4 are close to the upper limit of the sample rate, so the resulting frequencies might be aliased, leading to a linearly increasing pattern.

Second, unsteady flow modes are obtained by DMD. The Fourier modes reveal the location and frequency of the dominant flow structures; however, the spatial details are not sufficiently clear. Thus, DMD analysis is performed to obtain more detailed information. In the case of schlieren images, the main idea of the DMD method is as follows. Assume the schlieren images are saved as a set of pixels with dimension $n=p\times q$ in the time series $t=t_0, t_1,..., t_m$. The grayscale image snapshots, $x_t\in \mathbb{R}^{p\times q},~t=t_0,t_1,...,t_m$, can be described as a superposition of DMD modes as
\begin{equation}
\label{eq:superposition_dmd}
x_{t_j}=\sum_{i=1}^{r}b_i\phi_i e^{(\delta_i+i\omega_i)(j-1)\Delta t},~j=1,2,...,m,
\end{equation}
where $r$ is the maximum order of the mode after high-order truncation, $b_i$ is the mode amplitude, $\phi_i$ is the mode eigenvector, $\delta_i$ is the mode growth rate (defined as $\delta_i=\rm{Real}[\rm{log}(\mu_i)]/\Delta t$), $\omega_i$ is the mode angular frequency (defined by $\omega_i=\rm{Imag}[\rm{log}(\mu_i)]/\Delta t$), $\Delta t$ is the time interval, and $\mu_i$ is the eigenvalue of the $i$-th mode. These image snapshots can be rearranged into two sets, $\boldmath{X}=\{x_0,x_1,...,x_{m-1}\}\in \mathbb{R}^{n\times m}$ and $\boldmath{Y}=\{x_1,x_2,...,x_m\}\in \mathbb{R}^{n\times m}$, which have a constant time interval of $\Delta t=1/f_s$, where $f_s$ is the sampling rate. Given enough time, the lagged set $\boldmath{Y}$ is considered to be a linear mapping of the base set $\boldmath{X}$ \cite{schmid2010-dmd}. Thus, $\boldmath{Y}=\boldmath{AX}$. The flow field is then decomposed into DMD modes that are described by the eigenvectors and eigenvalues of the matrix $\boldmath{A}$. In this study, the streaming DMD (SDMD) algorithm \cite{hemati2014} is applied. The number of snapshots required to achieve the linear dependence between the two lagged datasets is investigated.
\begin{figure*}[!htb]
	\begin{minipage}[t]{1\linewidth}
		\centering
		\includegraphics[width=5.6in]{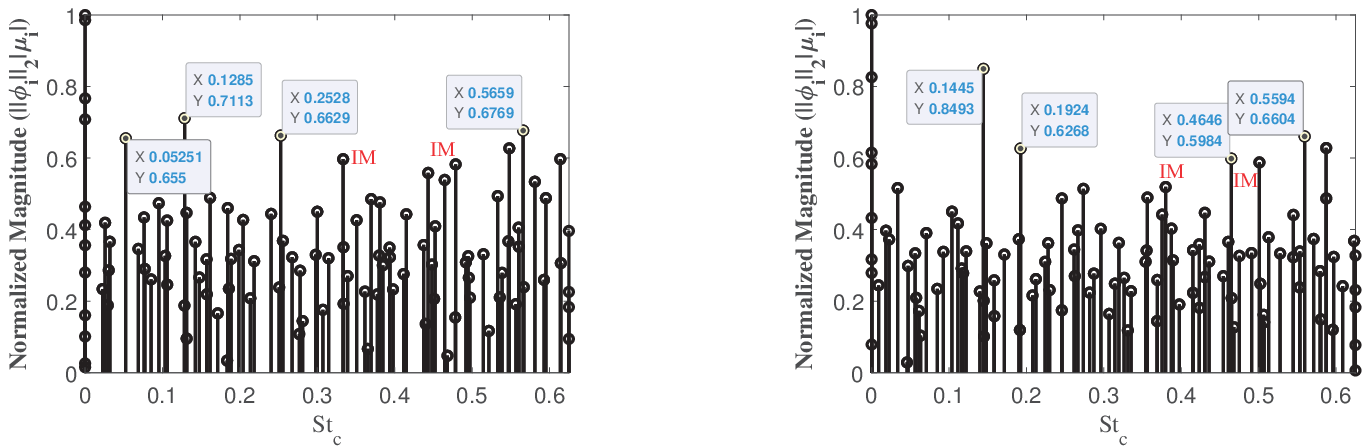}
	\end{minipage}
	\begin{minipage}[t]{1\linewidth}
		\centering
		\includegraphics[width=5.6in]{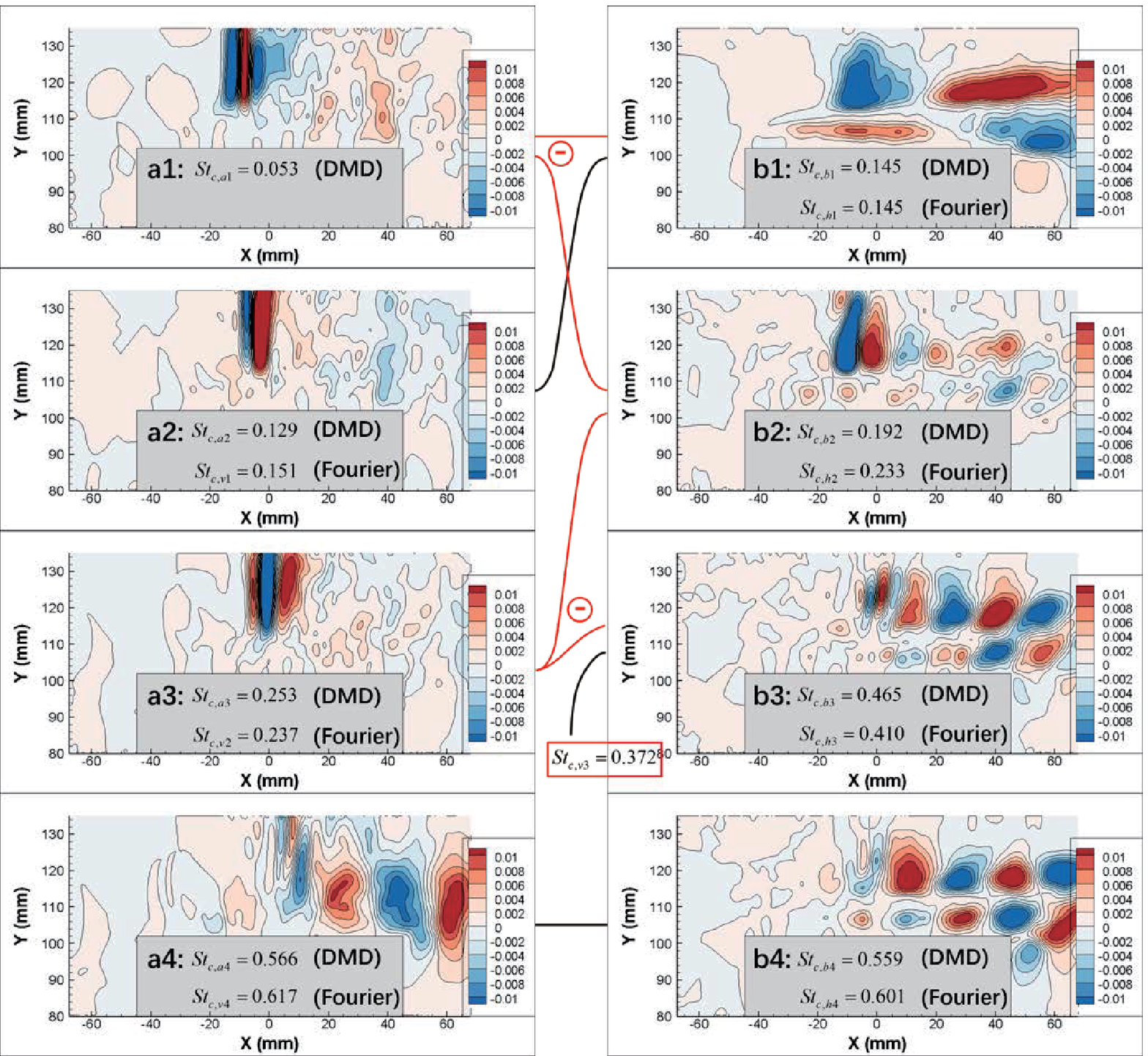}
		{\\\small(a) DMD of images from VL-cutoff.~~~~~~~~~~~~~~~~(b) DMD of images from HD-cutoff.}
	\end{minipage}
	\caption{\label{g:flat_dmd_spectra} DMD spectra and the related mode shapes at $r_p=0.297$ (PD120): (a) VL-cutoff; (b) HD-cutoff. Corresponding Fourier modes are attached. Coupling effects are illustrated by the linking lines. The symbol \circled{-} indicates a subtraction operation. Equivalence is indicated by black linking lines. Possible intermittent modes are labeled ``IM.''}
\end{figure*}
In most cases, a quasi-linear DMD mapping is achieved with about 2000 snapshots. The mode eigenvector is obtained after applying SDMD. A set of over-determined equations is then constructed from Eq.~(\ref{eq:superposition_dmd}), and the mode amplitude $b_i$ is solved by the least-squares method. Thereafter, the flow field of a single mode is reconstructed by $x_{t_j}=b_i\phi_ie^{(\delta_i+i\omega_i)(j-1)\Delta t},~j=1,2,...,m$. For each case, the DMD mode spectrum concerning the top 200 modes is calculated by the SDMD algorithm from 16\,000 snapshots. For each model, the mode spectra of the vertical and horizontal cutoffs are displayed, and the mode shapes of the dominant modes are given in terms of the mode spectra. The DMD amplitude is defined as the multiple of the norm of the eigenvector and the absolute value of the eigenvalue, $||\phi_i||_2 |\mu_i|$. This is normalized by the maximum value to reflect the overall and relative energy of the $i$-th mode. As a result, the mode amplitude is a measure of the total energy carried by a single mode structure across the distributed area. By contrast, the peak frequency of the Fourier transform is calculated locally. Therefore, the peak DMD frequencies may be different to those given by the Fourier transform, but the two are closely related.

As our focus is the transonic operating conditions, we present case PD120 ($r_p=0.297$). The left and right columns in Fig.~\ref{g:flat_dmd_spectra} represent the mode spectrum and mode patterns observed by the VL-cutoff and the HD-cutoff, respectively. An overall broadband feature is revealed, which implies that the tip clearance flow under transonic conditions is dominated both by periodic flow structures and fine-scale turbulent flows. Four relatively dominant modes are labeled under the broadband background. Based on the location and the mode shape, the first three modes in Fig.~\ref{g:flat_dmd_spectra}(a) are referred to as the SO modes related to the propagation and evolution of over-tip shock waves. See Figs.~\ref{g:flat_dmd_spectra}(a1)--\ref{g:flat_dmd_spectra}(a3) for their spatial mode shapes. The fourth mode shown in Fig.~\ref{g:flat_dmd_spectra}(a4) is related to the vortex-shedding and SLF modes. The shear-flapping modes are revealed by the labeled peaks in Fig.~\ref{g:flat_dmd_spectra}(b) with the mode shapes displayed in Figs.~\ref{g:flat_dmd_spectra}(b1)--\ref{g:flat_dmd_spectra}(b4). These modes are coupled with the vortex-shedding and SO modes. Similar to the locked-in effect observed for the Fourier modes, the coupling effect is revealed by the discrete pattern of the DMD SO modes, which should be broadband if the shock waves propagate in a free stream. This is also explained by the periodic locked-in oscillation of the throat size. The SLF effect leads to a dynamic change in the ``throat area," which controls the generation of unsteady shock waves. By examining the Strouhal numbers, the possible coupling relations are $St_{c,a1}\approx St_{c,b2}-St_{c,b1}$, $St_{c,a3}\approx St_{c,b3}-St_{c,b2}$, and $St_{c,a4}\approx St_{c,b4}$. Apart from the abovementioned modes, several narrowband peaks indicate possible intermittent modes, labeled ``IM.''

\subsection{Lock-in effect of over-tip shock waves}
\begin{figure}[!htb]
	\centering
	\includegraphics[width=3.in]{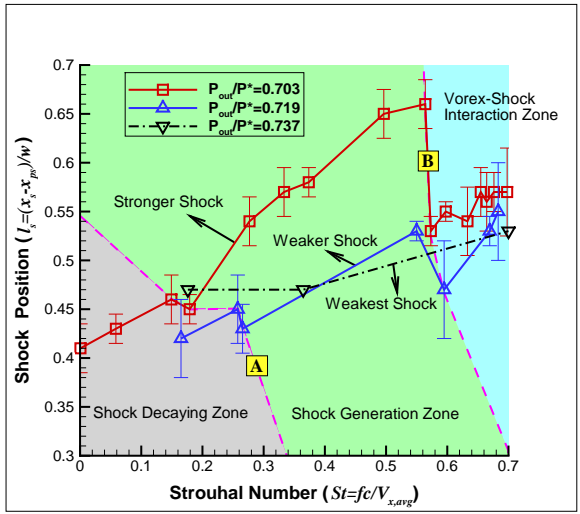}
	\caption{\label{g:frequency_vs_position} Shock spatial position changes with oscillation frequency. Positions are measured from Fourier modes, as shown by the subfigures in Fig.~\ref{g:spectra_along_line_flat_vl}. $x_s$,~$x_{ps}$, and $w$ are the $x$-coordinate of the center of the shock mode, the $x$-coordinate of the pressure side, and the width of tip clearance, respectively. Different color-flooded regions separated by purple dashed lines indicate three shock-motion zones. Error bars are based on the width of the Fourier mode patterns.}
\end{figure}

The generation and evolution of the upper-wall-bounded over-tip shock waves are revealed by the aforementioned Fourier and DMD modes (index \circled{3} is assigned to these flow structures). Measured from the Fourier mode patterns, the normalized over-tip shock positions are displayed with respect to the Strouhal number in Fig.~\ref{g:frequency_vs_position}. Three subzones can be identified. As labeled by the squared letters A and B, there is a clear boundary between the two zones if the over-tip shock waves are strong enough. This implies that not only the frequency, but also the position of the SO mode is locked-in by the SLF mode. This feature is essential for blade heat transfer, and is a key factor in control techniques such as casing-wall treatment and blade tip contouring.
Due to the zonal features of shock motions, the generation and evolution of over-tip shock waves are relatively discrete instead of continuous. This process is also revealed by the four-phase spatial translation of the DMD modes, as shown in Fig.~\ref{g:flat_dmd_spectra}(a). First, a dominant vortex-shedding mode is observed downstream of position $x=5.77$ mm with $St_c\approx0.566$. This mode is coupled with the shock generation by vortex/shear-layer/shock interactions. The shock waves are highly unsteady in this region. Thereafter, the shock waves propagate upstream, overtake one another, and merge with weaker shock waves, and the overall oscillation frequency decreases. As shown in Figs.~\ref{g:flat_dmd_spectra}(a)-a3, the dominant intermediate mode has a frequency of $St_c\approx0.253$, and the dominant mode at the final stage of the propagation and evolution is observed to be $St_c\approx0.053$. The intermediate modes should be broadband for shock waves propagation in a free stream. The strong discrete pattern is caused by the interaction with the vortex-shedding and SLF modes, which have almost-constant frequencies. After the final stage, as shown in Fig.~\ref{g:flat_dmd_spectra}(a)-a1, the shock waves start to fade out and remain spatially still near $x=-10$ mm ($38.5\% w$ from the PS edge).

\begin{figure}[!htb]
	\centering
	\includegraphics[width=3.in]{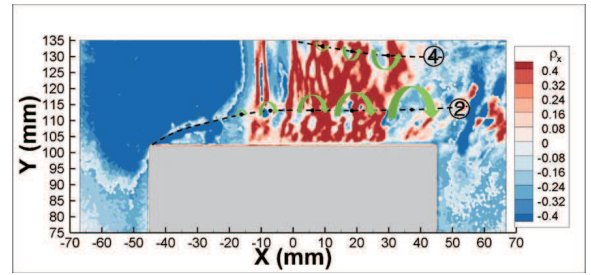}
	\caption{\label{g:shock_induced_vortex} Main vortex shedding \circled{2} and shock-induced vortices \circled{4}. The contour range is narrowed down to $[-0.4,+0.4]$ to enhance the clarity of the vortices.}
\end{figure}

The movement of shock waves along the upper casing wall induces boundary separation, resulting in shock-induced vortex shedding. Figure \ref{g:shock_induced_vortex} shows the main vortex shedding from the PS edge and the shock-induced vortices along the casing wall after the shock waves. The latter vortex shedding is assigned a flow structure index of \circled{4}.

\subsection{Identification of the escaping vortex-shedding mode}
The intermittency pattern of the third SO mode shown in Fig.~\ref{g:spectra_along_line_flat_vl} indicates that an intermittent flow structure emerges when the pressure difference is sufficiently strong. A similar phenomenon can be observed in Fig.~\ref{g:eigen_frequencies_vs_pds} as the mechanism switch at peak 3 and in Fig.~\ref{g:flat_dmd_spectra} as the narrowband and intermittent features labeled ``IM.'' This barely reported flow structure is revealed in detail by numerical analysis, and evidence can also be found in the schlieren images.

\begin{figure*}[!htb]
\begin{minipage}[t]{0.48\linewidth}
\centering
\includegraphics[width=2.8in]{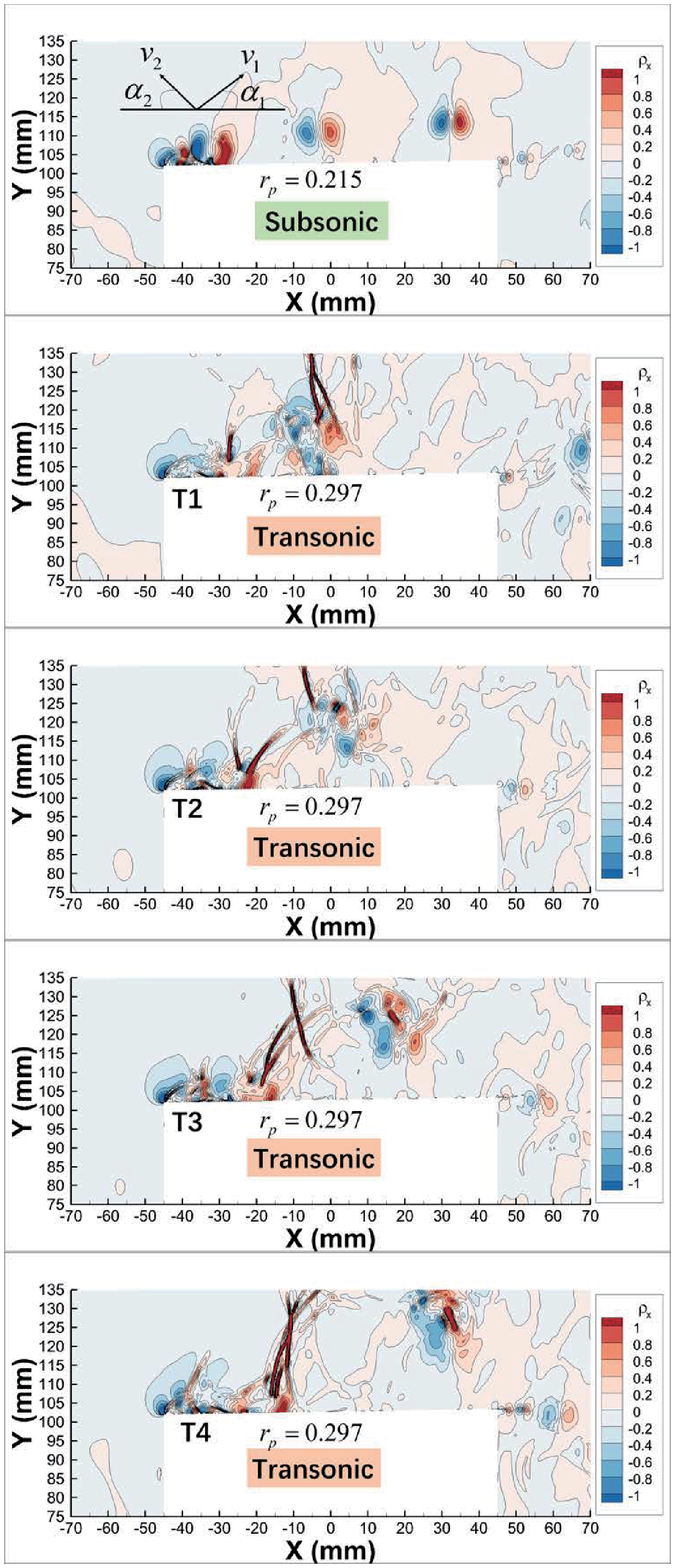}
{\\\small(a) $\rho_x$ predicted by IDDES.}
\end{minipage}
\begin{minipage}[t]{0.48\linewidth}
\centering
\includegraphics[width=2.8in]{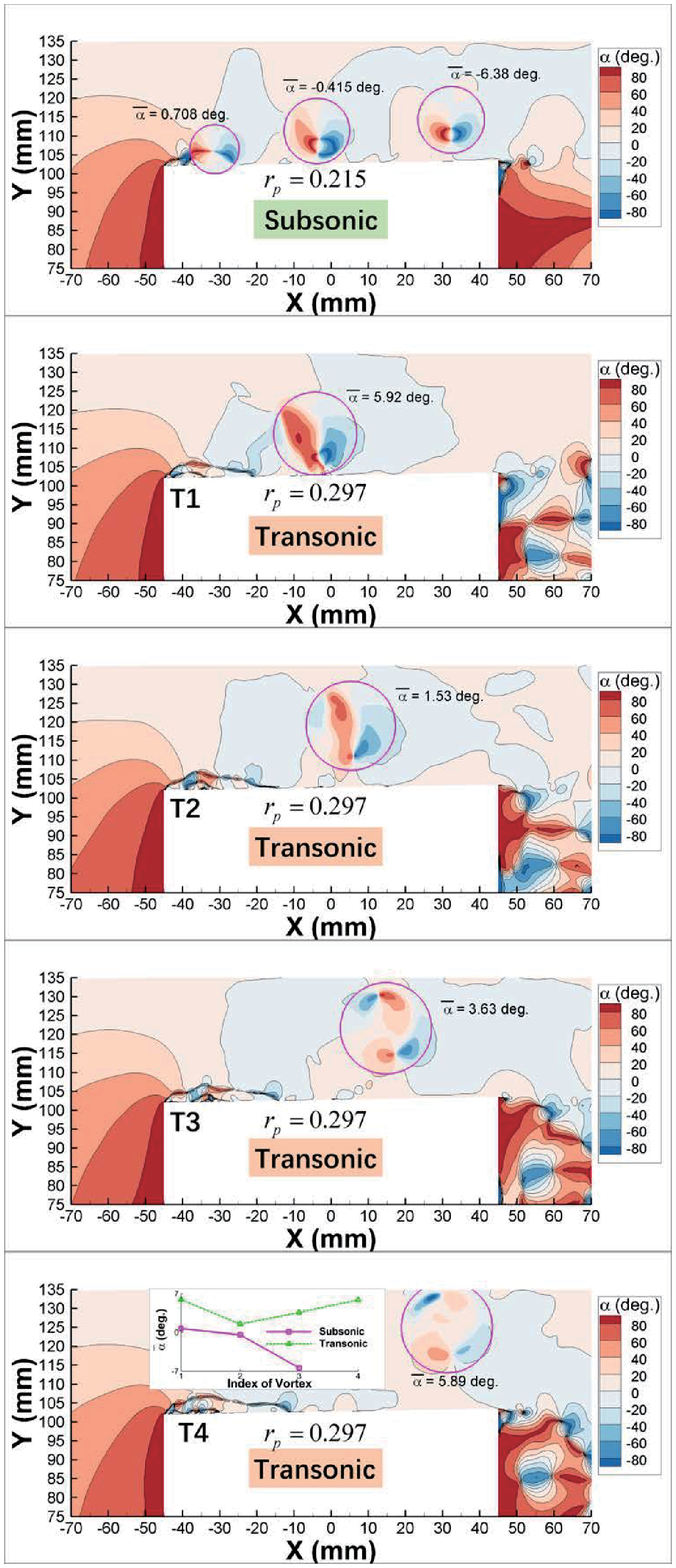}
{\\\small(b) Mean motion angle of the vortex core ($\bar{\alpha}$).}
\end{minipage}
\caption{\label{g:escaping_vortex} Identification of the escaping vortex mode: (a) snapshots of $\rho_x$ \textcolor{red}{(\href{https://pan.baidu.com/s/1QqoENNdCYsfX4DT-CUHUrg?pwd=acou}{Multimedia view})}; (b) flow angle $\alpha$ together with mass-weighted average of flow angle ($\bar{\alpha}$) \textcolor{red}{(\href{https://pan.baidu.com/s/1b1I1ElNU1v-fJEvZ3-M-7g?pwd=acou}{Multimedia view})}. Values of $\rho_x$ are normalized based on case PD120 ($r_p=0.297$). One snapshot for subsonic operating conditions ($r_p=0.215$) and a sequence of snapshots for transonic operating conditions ($r_p=0.297$). The variation of $\bar{\alpha}$ is attached to the last subfigure.}
\end{figure*}
\begin{figure*}[!htb]
	\begin{minipage}[t]{0.48\linewidth}
		\centering
		\includegraphics[width=2.8in]{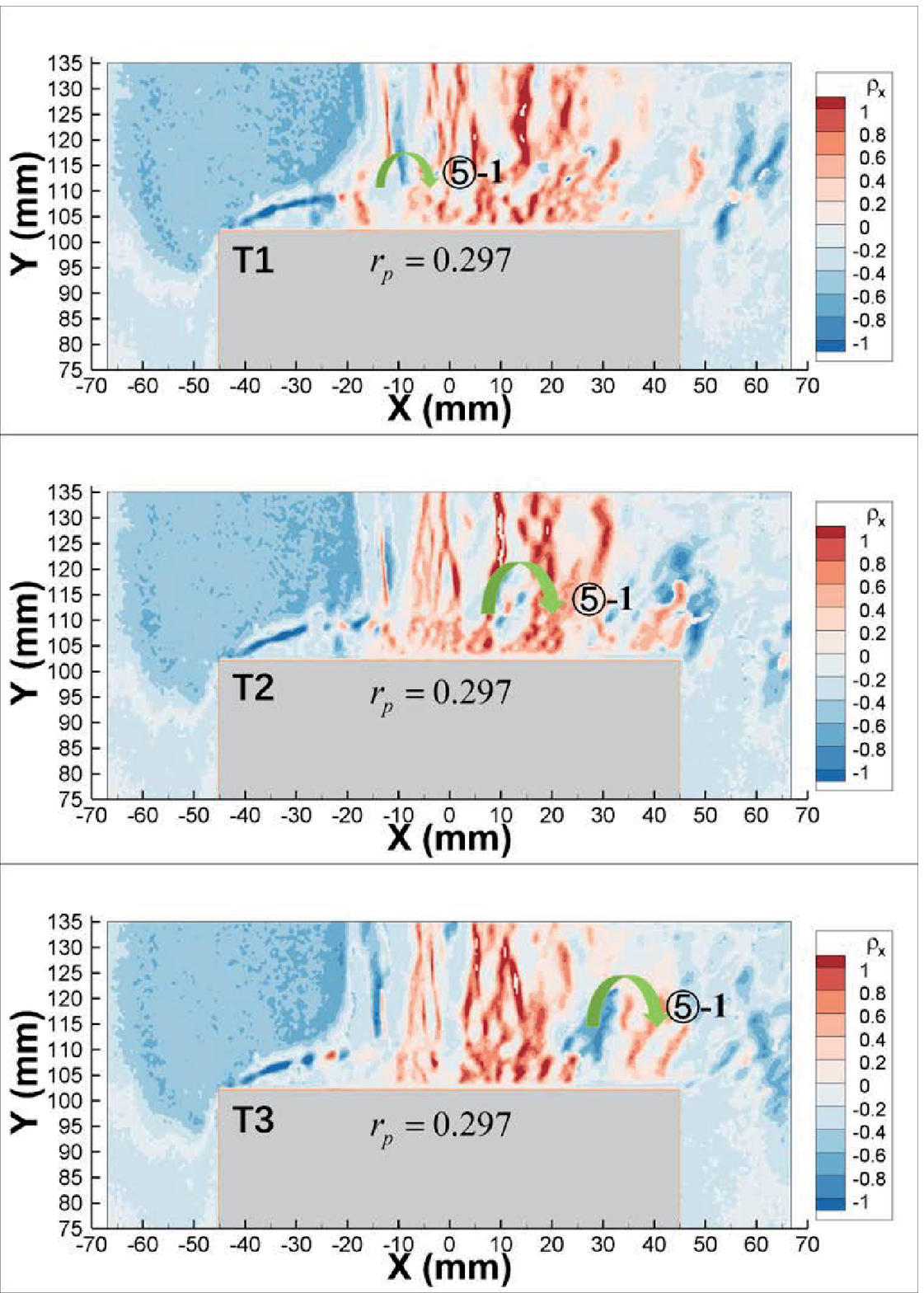}
		{\\\small(a) $\rho_x$ measured by vertical cutoff.}
	\end{minipage}
	\begin{minipage}[t]{0.48\linewidth}
		\centering
		\includegraphics[width=2.8in]{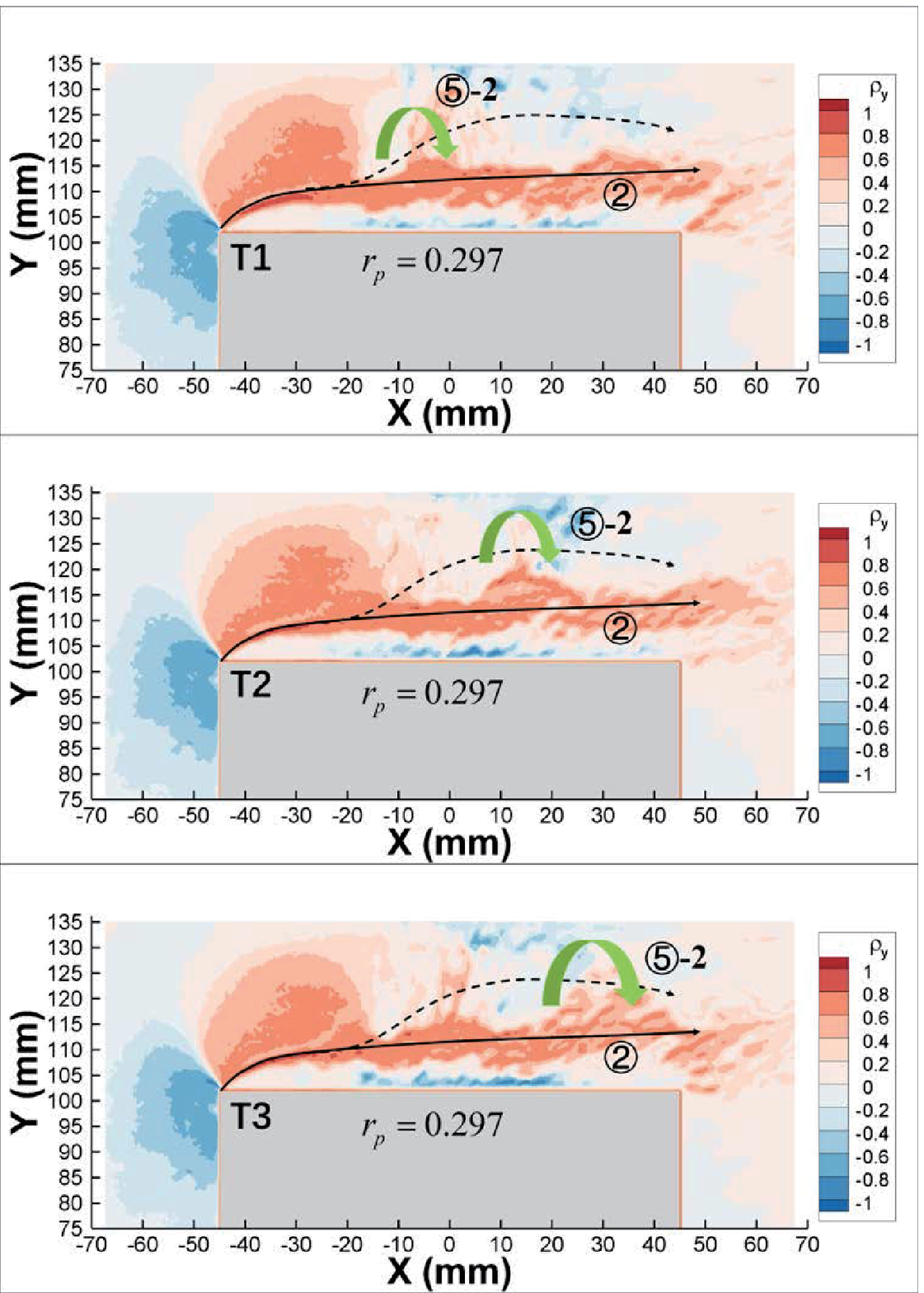}
		{\\\small(b) $\rho_y$ measured by horizontal cutoff.}
	\end{minipage}
	\caption{\label{g:escaping_vortex_vl_hd} Accumulated vortex shedding in case PD120 ($r_p=0.297$): (a) measured $\rho_x$; (b) measured $\rho_y$. \circled{5}-1: accumulated mode without escaping. \circled{5}-2: accumulated mode escaped from the vortex street.}
\end{figure*}
Figure \ref{g:escaping_vortex} shows two vortex-shedding patterns, and includes a snapshot in the case of subsonic flow and a four-step temporal sequence for the transonic flow case. The density gradient in the $x$-direction is displayed in the left column, as shown in Fig.~\ref{g:escaping_vortex}(a) \textcolor{red}{(\href{https://pan.baidu.com/s/1QqoENNdCYsfX4DT-CUHUrg?pwd=acou}{Multimedia view})}. The right column, Fig.~\ref{g:escaping_vortex}(b) \textcolor{red}{(\href{https://pan.baidu.com/s/1b1I1ElNU1v-fJEvZ3-M-7g?pwd=acou}{Multimedia view})}, shows the value of $\alpha$ and the mass-weighted averaged flow angle, which is defined as
\begin{equation}
\bar{\alpha}=\frac{\int_{\Omega} \rho \alpha dV}{\int_{\Omega} \rho dV},
\end{equation}
where $\Omega$ is the core region of the vortex that is defined by the vorticity threshold of $|\omega|\ge 10~s^{-1}$ and $V$ is the volume of fluid. As shown by the first subfigure in Fig.~\ref{g:escaping_vortex}(a) \textcolor{red}{(\href{https://pan.baidu.com/s/1QqoENNdCYsfX4DT-CUHUrg?pwd=acou}{Multimedia view})}, $\alpha$ is defined as the absolute angle between the velocity vector and the $x$-axis, that is,
\begin{equation}
\label{e:aov_definition}
\alpha=\begin{cases}
~~{\rm tan^{-1}}(v_y/v_x)(180^{\circ}/\pi),&if~v_x>0,\\
-{\rm tan^{-1}}(v_y/v_x)(180^{\circ}/\pi),&if~v_x<0.
\end{cases}
\end{equation}
Thus, $\alpha\in [-90^{\circ},+90^{\circ}]$. The vortex moves toward the casing wall if $\alpha>0$ and toward the blade tip if $\alpha<0$. As a result, $\bar{\alpha}$ represents the mean direction of movement of the air enclosed in the vortex. For subsonic flow, the trajectory of the PS-edge vortex shedding follows the vortex street. There exist 3--4 vortices in the displayed view, and this vortex number is relatively stable. The mean flow angle $\bar{\alpha}$ increases at the early stage before decreasing rapidly to bound the vortex in the region of the vortex street. This changes under high-loading operating conditions, when shock or strong compression waves emerge. The flow velocity after the shock wave is drastically reduced, and two key factors are modified as a result. First, the streamwise velocity of the vortex decreases significantly across the shock, which may lead to the accumulation and merging of 2--4 vortices. The strength of the vortex accumulation effect relies on the strength and distribution of the shock waves. Consequently, an intermittency pattern is achieved by the vortex-shedding mode. Second, the mean angle of the vortex, $\bar{\alpha}$, increases, which leads to the ``escape'' of accumulated vortices from the vortex street. This results in the escaping vortex-shedding mode, which is also intermittent and may interact with the casing wall boundary.

Some preliminary evidence of the vortex accumulation and escaping vortex-shedding mode is provided by the schlieren measurements. As shown in Fig.~\ref{g:escaping_vortex_vl_hd}(a), vortex structure \circled{5}-1 is much greater in size than average, and consists of fewer vortices. This is evidence of vortex accumulation resulting in vortex merging and flow intermittency. Thus, it is labeled as the accumulated vortex shedding without escaping from the vortex street. Vortex structure \circled{5}-2, as shown in Fig.~\ref{g:escaping_vortex_vl_hd}(b), escapes from the main vortex street. Therefore, it is labeled as the escaping vortex-shedding mode with the mode pattern given in Fig.~\ref{g:flat_dmd_spectra}(b3). All accumulated vortex-shedding modes that have an intermittency pattern are assigned a flow structure index of \circled{5}. This flow structure explains the intermittency of the third SO Fourier mode. The structure may interact with the upper wall and affect the generation of over-tip shock waves.

\subsection{Schematic drawing of flow structures in the clearance}
\begin{figure}[!b]
	\centering
	\includegraphics[width=3.4in]{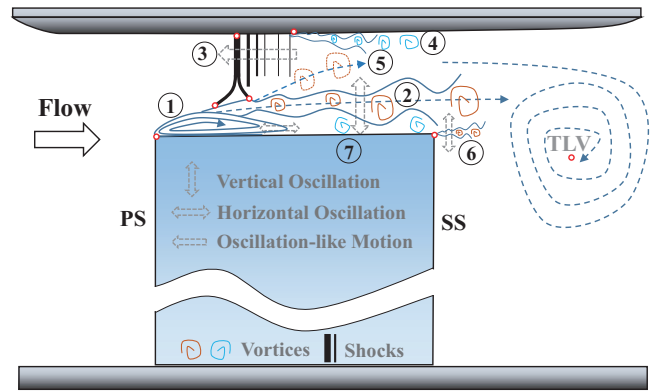}
	\caption{\label{g:flow_structures_flat}Schematic view of leakage flow structures in the gap region under transonic operating conditions: \circled{1} PS-edge separation bubble; \circled{2} main vortex street and shear layer; \circled{3} oscillating shock waves; \circled{4} shock-induced vortices; \circled{5} escaped PS-edge vortices from the vortex street; \circled{6} SS-edge vortex shedding; \circled{7} secondary vortices induced by the main vortex street.}
\end{figure}
\begin{figure*}[!htb]
	\begin{minipage}[t]{0.48\linewidth}
		\centering
		\includegraphics[width=2.8in]{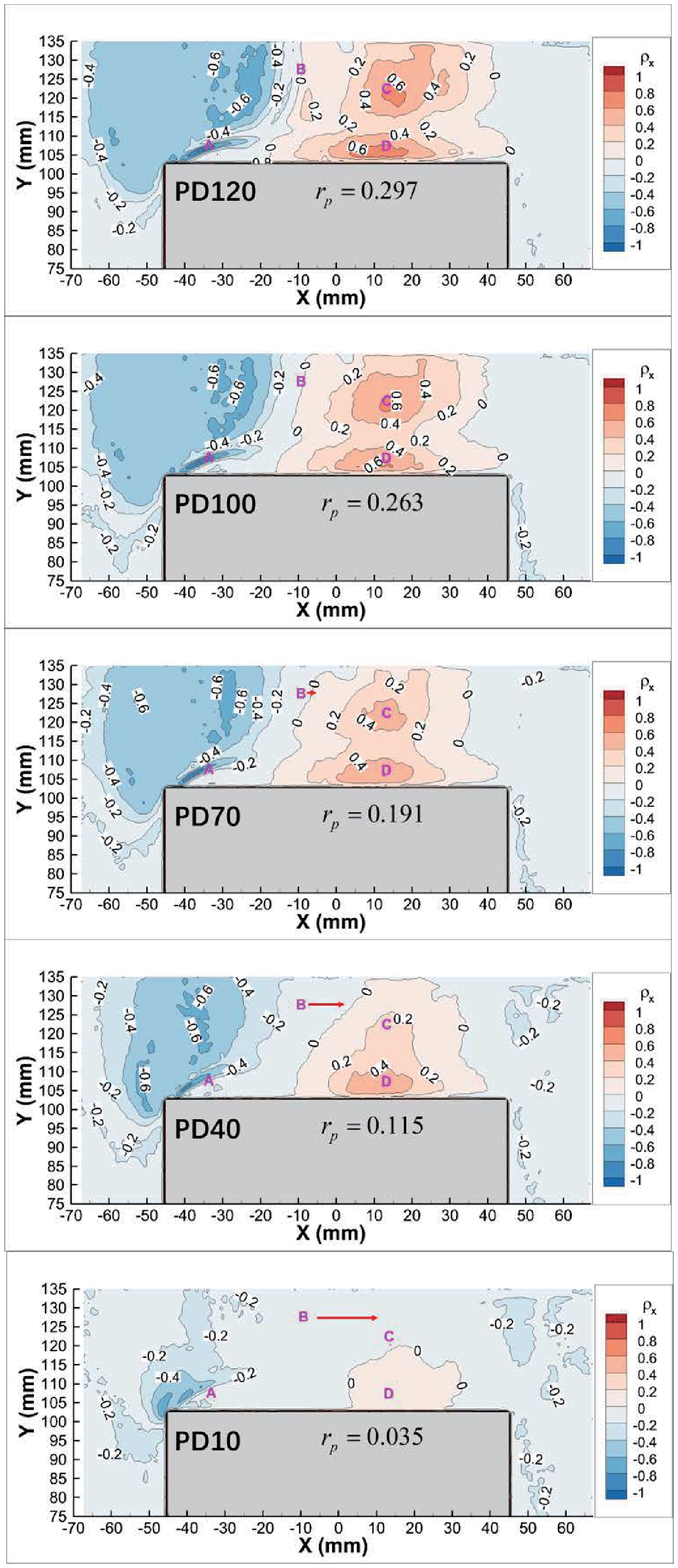}
		{\\\small(a) $\bar{\rho}_x$ measured by vertical cutoff.}
	\end{minipage}
	\begin{minipage}[t]{0.48\linewidth}
		\centering
		\includegraphics[width=2.8in]{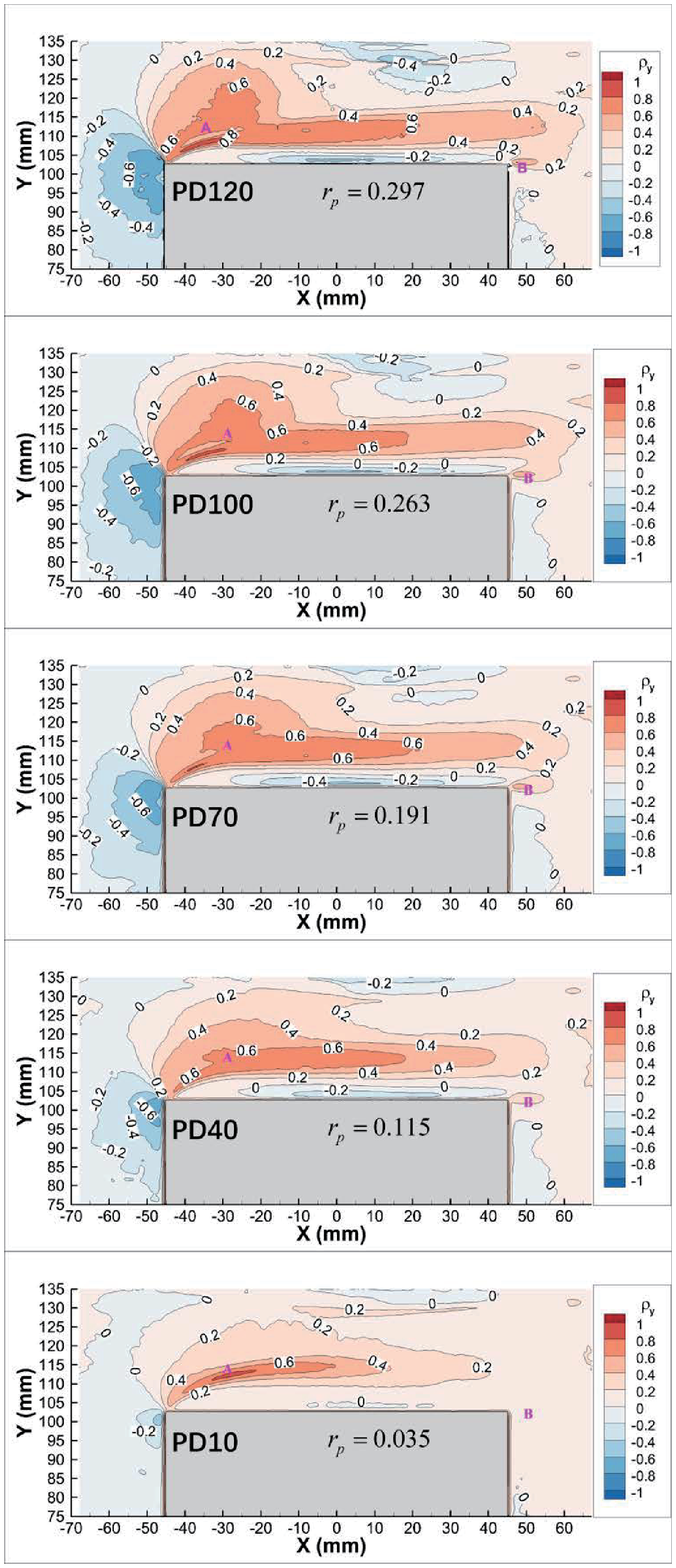}
		{\\\small(b) $\bar{\rho}_y$ measured by horizontal cutoff.}
	\end{minipage}
	\caption{\label{g:avg_image_flat_pds} Mean density gradient $(\bar{\rho}_x,\bar{\rho}_y)$ at five operating conditions: (a) measured $\bar{\rho}_x$; (b) measured $\bar{\rho}_y$. Maximum values are selected to normalize the density gradient based on the measurement in case PD120 ($r_p=0.297$). The vertical and horizontal cases are normalized independently.}
\end{figure*}
To summarize the aforementioned flow structures, a schematic drawing is shown in 
Fig.~\ref{g:flow_structures_flat}. Seven flow structures are discussed, with a focus on the transonic operating conditions. Each structure is assigned a circled index. For subsonic conditions, most flow structures remain, except for the SO mode, the shock-induced vortex shedding, and the escaping vortex shedding. Flow structures under supersonic conditions have been widely discussed in the literature. Figure~\ref{g:tip_flows_tips} provides an overview; see Wheeler \cite{wheeler2016} or Feng \cite{feng2021-pof} for further details.

\subsection{Averaged flow structures}
Given the unsteady flow patterns, temporal averaging provides a better understanding of the mean flow features. The mean density gradient $(\bar{\rho}_x,\bar{\rho}_y)$ is displayed in Fig.~\ref{g:avg_image_flat_pds}. Focusing on the case PD120 ($r_p=0.297$) in Fig.~\ref{g:avg_image_flat_pds}(a), the edge of the separation bubble (zone A) and the over-tip shock (zone B) are quantitatively represented. When entering the separation bubble, the density decreases rapidly, with the smallest density gradient observed at $(x,y)=(-35,106.4)$~mm. The flow changes from expansion to compression in zone B, which is the average upstream propagation limit of the unsteady shock waves. In the $x$-direction, the center of zone B is at about $x=-6.4$~mm, which is about $\Delta x=38.6$~mm ($42.9\%w$) from the pressure side. Zone C is the main compression zone, which generates most of the shock waves. The pressure recovery zone after the separation bubble is marked as zone D. Similarly, the overall field of $\bar{\rho}_y$ is displayed in Fig.~\ref{g:avg_image_flat_pds}(b). The flow structures highlighted here are the shear layers. The main shear layer, marked as A, originates from the edge of the pressure side and dominates the lower part of the tip clearance. In addition, a smaller shear layer, marked as B, is observed at the edge of the suction side where the tip leakage flow jets out of the clearance and interacts with the downstream edge. The main shear layer is relatively steady and thin initially. It thickens as the flow travels downstream due to the unsteady flapping and the self-thickening effect. The highest density gradient is observed at $(x,y)=(-36,107.9)$~mm, which is slightly different from the peak location of $\bar{\rho}_x$. As the pressure difference decreases, the regions of shock/compression waves, as indicated by letters B and C in Fig.~\ref{g:avg_image_flat_pds}(a), start to disappear. The reduced compression effect from the casing wall leads to larger separation bubbles, as indicated by letter A in Fig.~\ref{g:avg_image_flat_pds}(b). The flow is almost incompressible for cases PD10 ($r_p=0.035$) and PD20 ($r_p=0.063$).

\section{Conclusions}\label{s:conclude}
This paper has described an experimental investigation of a common generic blade tip model in a transonic wind tunnel as a means of studying the pressure-driven TLF. The flow fields were measured by pressure taps and time-resolved schlieren visualization. The distributions of static pressure and total pressure loss were reported along the tip surface. The length scale of the PS separation bubble, which is weakly affected by temperature, was observed to vary almost linearly with the pressure difference. A novel technique was applied to enhance the visualized flow details, with data from different schlieren cutoffs superposed. Image processing and additional numerical simulations revealed the dynamics of seven critical tip flow structures. The main conclusions from this study are as follows:

(1) Blade loading represented by the ratio of pressure difference, $r_p$, has strong effects on tip flow structures. First, an increase in compressibility occurs as the blade loading grows, resulting in the generation of over-tip shock waves. Second, under subsonic conditions, the trigger position of the shear-layer instability is monotonically delayed as blade loading increases; however, this pattern is reversed under transonic conditions. This feature implies that the PS flow acceleration, flow compression in the clearance region, and over-tip shock oscillation are critical factors in tip flow instabilities. 

(2) Under transonic conditions, several SO, SLF, and vortex-shedding modes were revealed by Fourier and DMD analysis. First, the SO modes were found to be locked-in with the SLF and vortex-shedding modes. Moreover, spatial analysis implies that both the frequency and position of the SO mode is locked-in by the shear-flapping modes. Second, the generation and evolution of over-tip shock waves were illustrated. Shock waves are generated in the compression region and propagate upstream. Moreover, shock overtaking occurs, resulting in a frequency reduction. Third, an intermittent mode was observed. Following this observation, an escaping vortex-shedding mode was discovered. Unsteady over-tip shock waves or strong compression waves are responsible for the accumulation and escaping of vortices. Based on the observed tip flow structures, a schematic drawing of TLF structures and related motions was proposed.

Additionally, to provide data for the validation of steady simulations, the mean flow field of the TLF was reported with respect to different operating conditions.


\begin{acknowledgements}
This work is supported by grants from the National Key Research and Development Project (Grant No.~2018YFA0703300), the National Natural Science Foundation of China (Grant No.~12102247), the Royal Society (Grant No.~IE131709), and the Key Laboratory of Aerodynamic Noise Control of China Aerodynamics Research and Development Center (Grant No.~ANCL20200202).
\end{acknowledgements}

\section*{Data availability}
The data that support the findings of this study are available
from the corresponding author upon reasonable request.

\end{CJK*}

\end{document}